\documentclass[manuscript]{aastex62}

\usepackage{newtxtext,newtxmath}

\usepackage[T1]{fontenc}
\usepackage{ae,aecompl}
\usepackage{pifont}

\usepackage{graphicx}  
\usepackage{amsmath}	
\usepackage{amssymb}	
\usepackage{longtable,booktabs}
\usepackage[countmax]{subfloat}

\graphicspath{{./}{figures/}}

\received{January 1, 2018}
\revised{January 7, 2018}
\accepted{\today}
\submitjournal{ApJS}

\shorttitle{Empirical relations for estimating stellar masses and radii}
\shortauthors{Moya et al.}

\begin{document}

\title{Empirical relations for the accurate estimation of stellar masses and radii \footnote{Released on March, 16th, 2018}}

\correspondingauthor{Andy Moya}
\email{A.Moya@bham.ac.uk}

\author{Andy Moya}
\affil{School of Physics and Astronomy, University of Birmingham, Edgbaston, Birmingham, B15 2TT, UK}
\affiliation{Stellar Astrophysics Centre, Department of Physics and Astronomy, Aarhus University, Ny Munkegade 120, DK-8000 Aarhus C, Denmark}

\author{Federico Zuccarino}
\affiliation{Universidad Internacional de Valencia (VIU), E-46021 Valencia, Spain}

\author{William J. Chaplin}
\affil{School of Physics and Astronomy, University of Birmingham, Edgbaston, Birmingham, B15 2TT, UK}
\affiliation{Stellar Astrophysics Centre, Department of Physics and Astronomy, Aarhus University, Ny Munkegade 120, DK-8000 Aarhus C, Denmark}

\author{Guy R. Davies}
\affil{School of Physics and Astronomy, University of Birmingham, Edgbaston, Birmingham, B15 2TT, UK}
\affiliation{Stellar Astrophysics Centre, Department of Physics and Astronomy, Aarhus University, Ny Munkegade 120, DK-8000 Aarhus C, Denmark}



\begin{abstract}

In this work, we have taken advantage of the most recent accurate stellar characterizations carried out using asteroseismology, eclipsing binaries and interferometry to evaluate a comprehensive set of empirical relations for the estimation of stellar masses and radii. We have gathered a total of 934 stars -- of which around two-thirds are on the Main Sequence -- that are characterized with different levels of precision, most of them having estimates of $M$, $R$, $T_{\rm eff}$, $L$, $g$, $\rho$ and [Fe/H]. We have deliberately used a heterogeneous sample (in terms of characterizing techniques and spectroscopic types) to reduce the influence of possible biases coming from the observation, reduction, and analysis methods used to obtain the stellar parameters. We have studied a total of 576 linear combinations of $T_{\rm eff}$, $L$, $g$, $\rho$ and [Fe/H] (and their logarithms) to be used as independent variables to estimate $M$ or $R$. We have used an error-in-variables linear regression algorithm to extract the relations and to ensure the fair treatment of the uncertainties. We present a total of 38 new or revised relations that have an adj-$R^2$ regression statistic higher than 0.85, and a relative accuracy and precision better than $10\%$ for almost all the cases. The relations cover almost all the possible combinations of observables, ensuring that, whatever list of observables is available, there is at least one relation for estimating the stellar mass and radius. 
\end{abstract}



\section{Introduction}

The existence of empirical relations among some observable stellar characteristics is well known from the initial works of \citet{Hertzsprung}, \citet{Russell}, and \citet{Eddington}. Improvements in the observational data, data analysis techniques and/or physical models have led to updates and revisions of these empirical relations \citep[see][for example]{Demircan_Kahraman}.

In recent years, a number of revisions of these empirical relations have been developed \citep{Torres10, Eker14, Gafeira12, Benedict16}. One of the common points of all these works is that they have used eclipsing binaries as observational targets.

Although some derived relations have been extensively used in the literature \citep[][for example]{Torres10}, the Mass-Luminosity relation and the Mass-Radius relation, two of the most conspicuous, have two main weak points: (i) The luminosity is, in general, known with great uncertainty, and this uncertainty is translated to the mass determination; and (ii) the radius is usually unknown.

Recent improvements in the observational data quality and quantity have opened new opportunities for re-evaluating these relations: 

\begin{itemize}
\item The first {\it Gaia} data release \citep{Gaia} has offered a new framework, providing accurate stellar luminosities on a significantly increased sample of stars. This has allowed a revision of the characteristics of some eclipsing binaries \citep{Stassun16}. Recently, the new Gaia DR2 \citep{DR2} has provided parallaxes with unprecedented precision.
\item The first group of accurate stellar radii obtained using interferometry from the VEGA optical interferometer at the CHARA array has been already published \citep{Ligi16}
\item Tens of accurate stellar densities have been confirmed thanks to planetary transits. \citet{Huber13}, for example, obtained stellar mean densities using asteroseismology and compared them with those derived from transits.
\item Hundreds of isolated stars have been characterized using asteroseismology yielding unprecedented precision, mainly thanks to {\it Kepler} \citep{Kepler} and {\it CoRoT} \citep{corot} data.
\end{itemize}

All these points together have opened a door to a complete revision of empirical relations for the accurate determination of stellar masses and radii.


In this paper we study all the possible empirical relations using the effective temperature ($T_{\rm eff}$), luminosity ($L$), surface gravity ($g$), mean density ($\rho$), and/or stellar metallicity ([Fe/H]) as independent variables and the stellar mass ($M$) or radius ($R$) as dependent variable. For this revision, we have gathered together data on all the stars in the literature that have been accurately characterized using asteroseismology, eclipses in detached binary systems, or interferometry.

As a result, 38 new or revised relations (18 for $M$ and 20 for $R$) are obtained with an adj-$R^2$ statistic larger than 0.85 (in fact, an 89$\%$ of them have a adj-$R^2 > 0.9$), an accuracy better than $10\%$ (except in three cases), and a precision better than $7.5\%$ (except in one case), depending on the observables available.

It is important to bear in mind that these relations are no substitute for the techniques that have been used to provide our source data. Our main aim is to condense the information provided by them into simple linear relations to estimate the stellar mass and radius, for use when source data are not available.

\section{Data sample}
\label{sec:data_sample}

We deliberately sought to build a calibration sample that is sufficiently extensive and heterogeneous as possible. Large samples more reliably reflect the population mean and also make it easier to identify and rule out outliers, and because it is also heterogeneous, the influence of possible biases inherent to the observations and data reduction and analysis methods used in obtaining the stellar parameters is significantly reduced.

In this study, we only considered studies based on asteroseismology, detached eclipsing binaries, and interferometry, since to date they are known to produce results with the highest precision. This high precision in the sample is critical for the reliability of the relations found. The main characteristics of these techniques from the point of view of our study are:

\begin{itemize}
\item {\it Asteroseismology:} Intrinsic stellar properties can be estimated from stellar pulsations. In the case we observe a large number of individual pulsational frequencies the stellar characterization can be enormously improved by fitting them to a grid of models \citep{Lebreton_Goupil}, generally using Bayesian methods \citep[and references therein]{Silva17}. One of the main uncertainties of the results coming from stellar model fitting is that they have a number of unknown free parameters and different physical descriptions for the same phenomena (opacities, nuclear reaction rates, EOS, etc.). The lack of sufficient observational constraints causes a large impact from these degrees of freedom in the final parameter estimation. In essence, asteroseismology covers this absence providing tens of additional observational constraints in the case where the individual frequencies are observed. These asteroseismic observational constraints are highly correlated, but the additional information provided by them is enough to obtain accurate characterizations. It is only recently, thanks to photometric observations made by {\it Kepler} and {\it CoRoT}, that we can obtain highly accurate oscillation frequencies mainly of solar-like stars \citep{Lund17, Davies16, Appourchaux12}. In addition, a set of non-seismic input constraints are required to guide the process, such as effective temperature and stellar metallicity, usually obtained from high-resolution spectroscopy. 


\item {\it Eclipsing binaries:} With high quality photometric and spectroscopic data we are able to derive, by means of dynamical effects, with high precision the intrinsic stellar properties of eclipsing binaries \citep{Andersen91, Torres10}. If the system can be solved spectroscopically, it is possible to obtain the mass from the radial velocity curves. Analyzing the light curves, which can be obtained photometrically, we can find the radius. The inclination and eccentricity of the orbit can be obtained using photometry and/or spectroscopy \citep{Bulut07}.  Moreover, if we measure the effective temperatures, we can then estimate the luminosities of the individual components \citep{Torres10}. Binary systems in close proximity are elongated towards each other because of mutual gravity; therefore, it is preferable to study only detached binaries where such effects are negligible \citep[see][]{Eker14}.

\item {\it Interferometry:} Currently it is not possible to resolve the angular diameter of stars with conventional telescopes.  This requires angular resolutions that are of the order of milliarcsec \citep{Boyajian13, Maestro13}.  However, optical interferometers offer spatial resolutions that are several orders of magnitude better than conventional telescopes.  The concept of interferometry is based on combining signals from an array of telescopes to obtain a unique interference pattern equivalent to a signal received by a single telescope with an aperture diameter equal to the maximum baseline of the array.  The interference pattern can be used to directly measure the angular diameter with remarkable accuracy, and when combined with the distance can be used to derive the radius. 
\end{itemize}

A thorough survey of recently published studies, based on the techniques previously described, produced an initial collection of close to 2000 entries. For each entry, the sample contains the following astronomical parameters: $M$, $R$, $T_{\rm eff}$, $L$, [Fe/H], $g$, and $\rho$, and their respective uncertainties. $M$, $R$, $g$, and $\rho$ were derived directly or indirectly from one of these three techniques. Stellar properties that were calculated, by definition, from already determined parameters (e.g. $g$ and $\rho$ obtained from the derived $M$ and $R$, and not from the observational data) were not taken into account if possible. [Fe/H] was obtained mainly from spectroscopy. $T_{\rm eff}$, depending on the case, comes from spectroscopic or photometric observations\footnote{In the case of one of the main sources of our sample, \citet{Serenelli17}, we used $T_{\rm eff}$ obtained from photometry, following the advice of the authors (A. Serenelli, private communication).}. $T_{\rm eff}$ and [Fe/H] are also necessary inputs for deriving some of the other parameters, e.g. as constraints for asteroseismic grid-based modeling. 

Some studies provide $L$, but we have also obtained the bolometric luminosity using the VO Spectral Energy Distribution Analyzer \citep[VOSA\footnote{http://svo2.cab.inta-csic.es/theory/vosa/index.php}$^{,}$\footnote{Note that VOSA is not recommended to obtain luminosities of stars that are in binary systems.};][]{2008AA...492..277B}.

VOSA queries tens of photometric catalogs accessible through VO services, builds the spectral energy distribution (SED), compares the observed SED to the synthetic photometry from several theoretical models, and computes the estimated $L_{\rm SED}$ for each star. For this study, we chose the Kurucz ODFNEW/NOVER theoretical model for stellar atmospheres \citep{1997AA...318..841C}. 

By adding the extinction parameter $A_{\rm v}$ \citep[taken from][]{Huber16, Mathur17} to the observed photometry, we compensate the effects of interstellar reddening and help in the correct estimation of the SED shape. Subsequently, to estimate the total flux, VOSA finds the best fit of the SED photometric points approximating a value of $T_{\rm eff}$ in fixed steps of 250K. Wanting to be conservative about the error in $L_{\rm SED}$, we chose a fixed error in $T_{\rm eff}$ equal to half the grid step, or 125K, for all estimated $T_{\rm eff}$ values. 

The transformation from total flux to bolometric luminosity was achieved utilizing distances from the Gaia DR2\footnote{https://www.cosmos.esa.int/web/gaia/dr2} \citep{DR2}, with an unprecedented accuracy of less than 1\%. 

The combination of all these elements allowed us to work with errors in $L_{\rm SED}$ that are in a range of 8-12\%.

The consolidated sample contains, as was to be expected, duplicate observations for the same stars; therefore, the first step was to filter the sample to remove duplicate entries. The choice was made based mainly on precision, giving priority to observations with errors of less than 7\% in $M$, $R$, and $T_{\rm eff}$, and of less than 10\% in $L$. The selection process prioritizes, first, observations that comply with the imposed thresholds in all 4 parameters, second, in 3 parameters, and so on. If after completing the selection process there are still duplicate entries, the tie is resolved by selecting by catalog \citep[i.e.][etc., in an intended order.]{Serenelli17, Silva17}, following intra-technique heterogeneity and reliability criteria. 

Not surprisingly, we found that not all catalogs provided data for all parameters. [Fe/H] is not available for all observations in the sample, and the same is true for $g$ or $\rho$. In Table \ref{tab:tab1} we summarize the contributing stellar parameters by catalog, and identify the corresponding units of measurement.

\begin{figure}
 \includegraphics[width=\columnwidth]{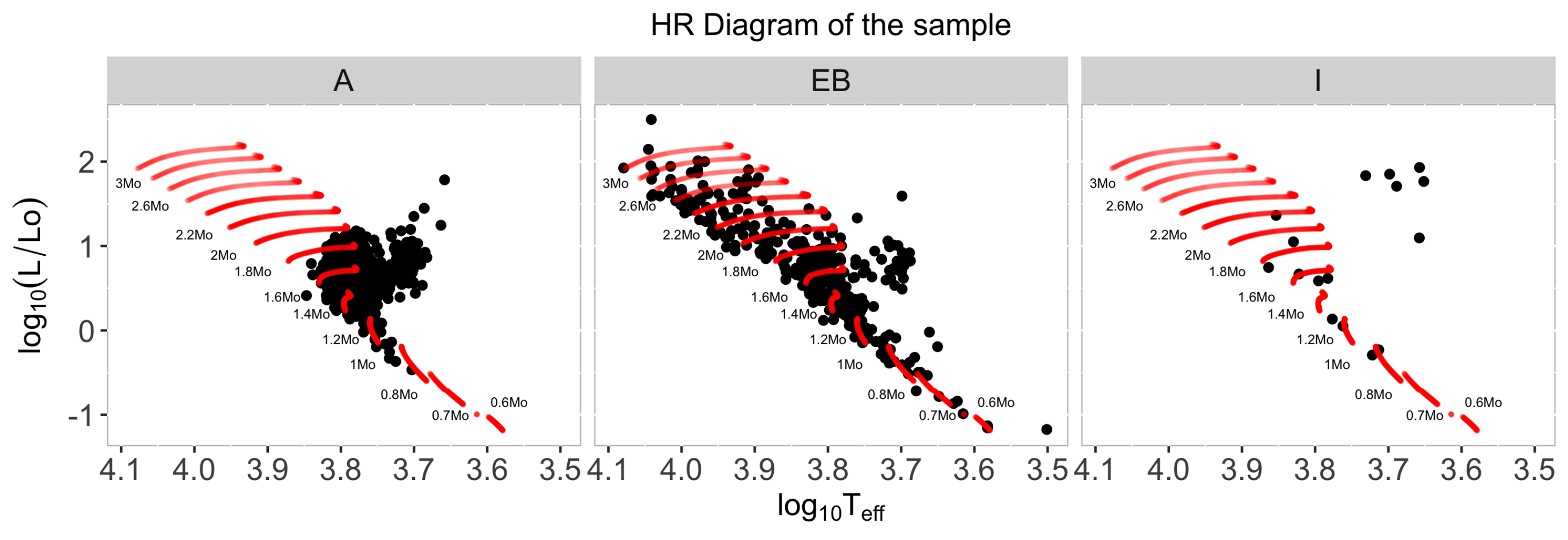}
 \caption{HR-diagram with the 934 MS/post-MS stars that make up the final calibration sample. The aligned red points are theoretical model tracks of different masses obtained using PARSEC \citep{2012MNRAS.427..127B}. Each panel accounts for the techniques used for studing the star: A = Asteroseismology, EB = Eclipsing Binaries, I = Interferometry.}
 \label{fig:HR}
\end{figure}

\begin{table}
	\centering
	\caption{Summary of the stellar parameters that each catalog contributes to the final calibration sample. The units of measurement corresponding to each parameter are also identified. }
    \label{tab:tab1}
	\begin{tabular}{lcccccccc} 
		\hline
		Catalog & $M$ & $R$ & $T_{\rm eff}$ & $L$ & [Fe/H] & $g$ & $\rho$ \\
                & [$M_\odot$] & [$R_\odot$] & K & [$L_\odot$] & [dex] & [cm/s$^2$] & [$\rho_\odot$] \\
		\hline
		Chaplin'14 & \ding{51} & \ding{51} & \ding{51} & $L_{\rm SED}$ & \ding{51} & \ding{51} & \ding{51} \\
        Eker'14 & \ding{51} & \ding{51} & \ding{51} & $L$ & some & \ding{51} & \ding{53} \\
        Huber'13 & \ding{51} & \ding{51} & \ding{51} & $L_{\rm SED}$ & \ding{51} & \ding{53} & \ding{51} \\
        Ligi'16 & \ding{51} & \ding{51} & \ding{51} & $L$ & \ding{51} & \ding{53} & \ding{53} \\
        Lund'16 & \ding{51} & \ding{51} & \ding{51} & $L_{\rm SED}$ & \ding{51} & \ding{51} & \ding{51} \\
        Malkov'07 & \ding{51} & \ding{51} & \ding{51} & $L$ & some & \ding{53} & \ding{53} \\
        Serenelli'17 & \ding{51} & \ding{51} & \ding{51} & $L_{\rm SED}$ & \ding{51} & \ding{51} & \ding{51} \\
        Silva'15 & \ding{51} & \ding{51} & \ding{51} & $L$ & \ding{51} & \ding{51} & \ding{51} \\
        Silva'17 & \ding{51} & \ding{51} & \ding{51} & $L$ & \ding{51} & \ding{51} & \ding{51} \\
        Torres'10 & \ding{51} & \ding{51} & \ding{51} & $L$ & some & \ding{51} & \ding{53} \\
        Welsh'12 & \ding{51} & \ding{51} & \ding{51} & $L$ & \ding{51} & \ding{51} & \ding{53} \\
		\hline
	\end{tabular}
\end{table}

The final calibration sample consists of 934 stars, of which 726 are on the Main-Sequence (MS) and 208 are post-Main-Sequence (post-MS) Subgiants or Giants. The most significant contributions come from \citet{Eker14} with 222 stars, and \citet{Serenelli17} with 397 stars. The MS/post-MS classification was done using the evolutionary tracks described in \citet{Rodrigues17}, with solar metallicity. The impact of this classification in our results, when tracks with other characteristics are used, is analyzed in section \ref{sec:MS}. The sample contains stars from a wide range of spectral types, but the vast majority, or more than 700, are of types F or G. In Fig. \ref{fig:HR} we show the location of the MS/post-MS stars in the HR-diagram.  We also show some theoretical model tracks as reference obtained using PARSEC \citep{2012MNRAS.427..127B}.

Follows a brief overview of the articles/catalogs that were used as input to build the final calibration sample (see Table \ref{tab:tab2} for reference).

\begin{table}
	\centering
	\caption{Number of MS and post-MS stars per catalog that ended up in the final calibration sample, after filtering out duplicates. The main detection technique is also indicated.}
	\label{tab:tab2}
	\begin{tabular}{lccc} 
		\hline
		Catalog & Method & MS & post-MS\\
		\hline
		Chaplin'14 & Asteroseismology & 72 & 32\\
        Eker'14 & Binaries & 204 & 18\\
        Huber'13 & Asteroseismology & 24 & 19\\
        Ligi'16 & Interferometry & 10 & 6\\
        Lund'16 & Asteroseismology & 28 & 5\\
        Malkov'07 & Binaries & 34 & 1\\
        Serenelli'17 & Asteroseismology & 275 & 122\\
        Silva'15 & Asteroseismology & 29 & 0\\
        Silva'17 & Asteroseismology & 22 & 0\\
        Torres'10 & Binaries & 24 & 5\\
        Welsh'12 & Binaries & 4 & 0\\
		\hline
	\end{tabular}
\end{table}

\citet{2014ApJS..210....1C}, using asteroseismic analysis based on {\it Kepler} photometry of the first 10 months of science operations, determined $M$ and $R$ of more than 500 stars. The study can be divided into 2 subsets. A subset of 87 stars with atmospheric properties ($T_{\rm eff}$ and [Fe/H]) obtained by high-resolution spectroscopy \citep[see][]{2012MNRAS.423..122B}. The spectra were obtained with the ESPaDOnS spectrometer at the 3.6m CFHT telescope and with the NARVAL spectrometer at the 2m Bernard Lyon telescope. And a subset of 416 stars with $T_{\rm eff}$ obtained from complementary photometry (we used the SDSS-calibrated values). The authors adopted a fixed [Fe/H] value for the grid-based modeling, corresponding to an average value for the field\footnote{In the sample, we replaced the fixed [Fe/H] value with more recent and more accurate [Fe/H] values taken from \citet{Serenelli17} and the KIC catalog by \citet{Mathur17}.}. $M$, $R$, $g$, and $\rho$ were determined combining the Bellaterra Stellar Properties Pipeline \citep[BeSPP;][]{2013MNRAS.429.3645S} with the Garching Stellar Evolution Code \citep[GARSTEC;][]{2008Ap&SS.316...99W}.  The authors did not estimate $L$, so $L_{\rm SED}$ from VOSA was used instead.

The \citet{Eker14} catalog consists of 257 double line spectroscopic eclipsing binaries that are detached. The catalog contains $M$, $R$, $T_{\rm eff}$, $L$, and $g$. The complete catalog consists of stars with effective temperatures ranging from 2750K to 43000 K, but for our sample, we took into account mainly AFGK stars. The catalog itself is a compilation of multiple studies of light curves and radial velocities that can be found in the literature. In total, it contributes 222 stars to our sample, of which more than 90\% have an error in $M$, $R$, and $T_{\rm eff}$ of less than 7\%, and around 60\% have an error in $L$ of less than 10\%. The authors did not provide [Fe/H] and $\rho$. We completed the sample with some metallicities taken from the DEBCat catalog of \citet{2015ASPC..496..164S}. 

\citet{Serenelli17} present an asteroseismic analysis of 415 stars observed by {\it Kepler}. The authors provide two sets of data based on two independent $T_{\rm eff}$ scales. They favor the data derived from photometry in the SDSS {\it griz} bands, and those are the data we selected for our sample.  $M$, $R$, $g$, and $\rho$ were determined by grid-based modeling using a combination of BeSPP and GARSTEC. The catalog was completed with $L_{\rm SED}$ from VOSA, with a mean error of around 9\%; thanks to the small uncertainties of the Gaia DR2 survey. [Fe/H] comes from spectroscopic observations from the APOGEE survey \citep{2017AJ....154...94M}.

\citet{2016PASP..128l4204L} present an asteroseismic analysis of 33 solar-like stars observed by the K2 mission. The modeling of $M$, $R$, and $\rho$ was done through grid-based modeling using the Bayesian Stellar Algorithm \citep[BASTA;][]{2015MNRAS.452.2127S} and GARSTEC. $T_{\rm eff}$, [Fe/H], and $g$ were derived using the Stellar Parameter Classification pipeline \citep[SPC;][]{2012Natur.486..375B} from spectra obtained with the TRES spectrometer at the 1.5m Tillinghast telescope. The authors did not provide $L$, so we used $L_{\rm SED}$ instead.

The oscillation frequencies for the asteroseismic analysis of the \citet{Huber13} catalog were acquired by {\it Kepler} photometry. $T_{\rm eff}$ and [Fe/H] were obtained from spectroscopic observations using four different instruments installed in terrestrial observatories: the HIRES spectrometer at the 10m Keck telescope, the FIES spectrometer at the 2.5m NOT telescope, the TRES spectrometer at the 1.5m Tillinghast telescope, and the Tull-Coud\'e spectrometer at the 2.7m Harlan J. Smith telescope. The authors used different model pipelines (ASTEC, BaSTI, Padova, Yonsei-Yale, among others) to compute a likelihood function to determine the best fitting-model with which they estimated $M$ and $R$ of 77 stars (all confirmed or candidate planet-hosting stars). Luminosities come from VOSA, and $\rho$ comes from scaling relations. They did not provide $g$.

The catalog of \citet{2015MNRAS.452.2127S} is a subset of 33 stars of \citet{Huber13}, with some differences. $T_{\rm eff}$ and [Fe/H] were obtained by high resolution spectroscopy. $M$, $R$, $g$, and $\rho$ were determined using BASTA with a GARSTEC grid. $L$ was derived applying the Infrared Flux Method \citep[IRFM;][]{2010AA...512A..54C} to {\it griz} band photometry (and other sources). Data in this catalog is notably precise; all stars in the sample display errors in $L$ of less than 10\%, and less than 7\% in $M$ and $R$.  

\citet{Silva17}, using several model pipelines (AIMS, ASTFIT, BASTA, among others), present the stellar properties ($M$, $R$, $L$, $g$, and $\rho$) of 66 solar-like stars with what is the best asteroseismic data to date. The sample is known as the {\it Kepler} LEGACY sample. The authors emphasize that in general there is an excellent level of agreement between the different models, and for our sample, we selected the data that come from using BASTA with the GARSTEC grid. $T_{\rm eff}$ and [Fe/H], used as constraints, come from various sources in the literature \citep[][and others]{2012ApJS..199...30P, 2014ApJ...787..110C, 2015ApJ...808..187B}. 

\citet{Torres10} contribute with a catalog of 95 pairs of detached eclipsing binaries with errors in $M$ and $R$ of less than 3\%. The authors provide $M$, $R$, $T_{\rm eff}$, $L$, [Fe/H] (some, when available), and $g$. Their work supersedes and more than doubles an earlier work done by \citet{Andersen91}, and as in the case of \citet{Eker14}, the data comes from multiple sources in the literature \citep[][among others]{2000AJ....119.1942T, 2004AJ....128.1340L, 2008AA...487.1095C}. 

\citet{2012Natur.481..475W} present the stellar properties ($M$, $R$, $T_{\rm eff}$, $L$, [Fe/H], and $g$) of two pairs of sun-like stars with a circumbinary low-density gas giant planet orbiting each pair. The stars are eclipsing binaries and the stellar properties were obtained by combining {\it Kepler} photometry with spectroscopy using the HRS spectrometer at the Hobby-Eberly 10m telescope, the Tull-Coud\'e spectrometer at the 2.7m Harlan J. Smith telescope, the FIES spectrometer at the 2.5m NOT telescope, and the HIRES spectrometer at the 10m Keck telescope. Errors in $M$ and $R$ are less than 1\%. The study did not provide an estimate of $\rho$. All 4 stars are on the Main-Sequence.

\citet{Ligi16}, using the {\it VEGA} optical interferometer at the {\it CHARA} array, determined the angular diameter of 18 stars and estimated $R$ with an error of less than 5\%. By fitting spectral energy distributions taken from the VizieR SED viewer, they determined the total flux and derived $L$ and $T_{\rm eff}$. $M$ was derived using the Padova and Trieste Stellar Evolution Code \citep[PARSEC;][]{2012MNRAS.427..127B}.  [Fe/H], used as a constraint, was compiled from the literature.  All stars in the sample display errors in $L$ of less than 10\%. The stars are also potential exoplanet host stars.

\citet{Karovicova18} found that the angular diameters they have derived of 3 metal-poor benchmark stars are smaller than those derived by other interferometric studies of the same stars \citep{Creevey12, Creevey15}. They claim that comparative data between photometric and interferometric $T_{\rm eff}$ suggest that diameters of less than 1 milliarcsec appear to be systematically larger than expected. They argue the difference is due to calibration errors, and that the discrepancy tends to increase with the decrease in angular diameter. All but 3 stars of \citet{Ligi16} have angular diameters of less than 1 milliarcsec. \citet{Karovicova18} suggest that the \citet{Ligi16} catalog could be overestimating $R$. In any case, this subsample is always a small percentage of the total sample.

The \citet{2007MNRAS.382.1073M} catalog is based on a set of detached main-sequence double-lined eclipsing binaries. The catalog is a collection of studies found in the literature, the vast majority from the 1990s and early 2000s \citep[][and others]{1993BICDS..42...27M}, and compiles $M$, $R$, $T_{\rm eff}$, and $L$ of 215 stars. We chose a subset of stars that are mainly AFG; with a mean error in $M$, $R$, and $T_{\rm eff}$ of about 3\%, and 12\% in the case of $L$. The study did not come with [Fe/H], $g$, and $\rho$.

Lastly, since [Fe/H] is sometimes determined with high uncertainty, we augmented our sample by adopting metallicities from the California Kepler Survey \citep[published by][]{2017AJ....154..107P}. The catalog works with high-resolution spectra obtained with the HIRES spectrometer at the 10m Keck telescope. The error was reduced (by 16\%, on average) for 83 stars, all from the asteroseismology sample.

\section{Data analysis}

To analyze the data, we followed a three-step procedure. We first defined the combinations of variables to be tested, then we selected the best subset of stars for analyzing this particular combination, and finally we applied a Generalized Least Squares with Measurement Error  \citep[GLSME, see Section \ref{sec:analysis_method}]{Hansen12} algorithm to obtain the regression coefficients, their errors, and some statistics to analyze the quality of the regression (adj-R-Squared statistic, from now on $R^2$ for simplicity; mean accuracy, Acc; and mean precision, Prec, of which more below. See, for example, \citet{Fuller08}).
%
%
%
%

\subsection{Combinations of variables}
\label{sec:comb_var}

One of the main aims of this work is to study all the empirical relations possible for estimating stellar masses and radii, selecting those providing a better description of the data.  We have searched for any possible combination describing the information contained in the data, no matter which variables are combined with others.

\begin{table}
\centering
\caption{Percentage of stars in the global sample characterized by each technique. \label{table:percentages_tech}
}
\begin{tabular}{cccc}
  \hline
Technique & Astero. & Eclips. Bin. & Interf.\\ 
  \hline
$\%$ & 67.24 & 31.05 & 1.71\\ 
   \hline
\end{tabular}
\end{table}

\begin{table}
\centering
\caption{Percentage of stars in the global sample by their spectral types. \label{table:percentages_type}
}
\begin{tabular}{ccccccc}
  \hline
Spect. Type & B & A & F & G & K & M\\ 
  \hline
$\%$ & 1.9 & 9.1 & 48.9 & 29.8 & 10.2 & 0.1\\ 
   \hline
\end{tabular}
\end{table}


In addition, we have also allowed combinations where variables are substituted by their logarithms. That is, we have studied all possible combinations with the form:
 \begin{eqnarray}
 M\;\mathrm{or}\;R\;\mathrm{or}\;{\rm logM}\;\mathrm{or}\;{\rm logR} = & f(T_{\rm  eff}\;\mathrm{or}\;{\rm log}T_{\rm eff},\,L\;\mathrm{or}\;{\rm logL},\nonumber \\ 
&\,g\;\mathrm{or}\;{\rm log}g,\,\rho\;\mathrm{or}\;{\rm log}\rho,\,{\rm [Fe/H]})
 \label{eq:gen_eq}
 \end{eqnarray}
Combinations of one single variable, two, three, four, and five variables are allowed. This means a total of 576 possible combinations.

\begin{figure}
 \includegraphics[width=\columnwidth]{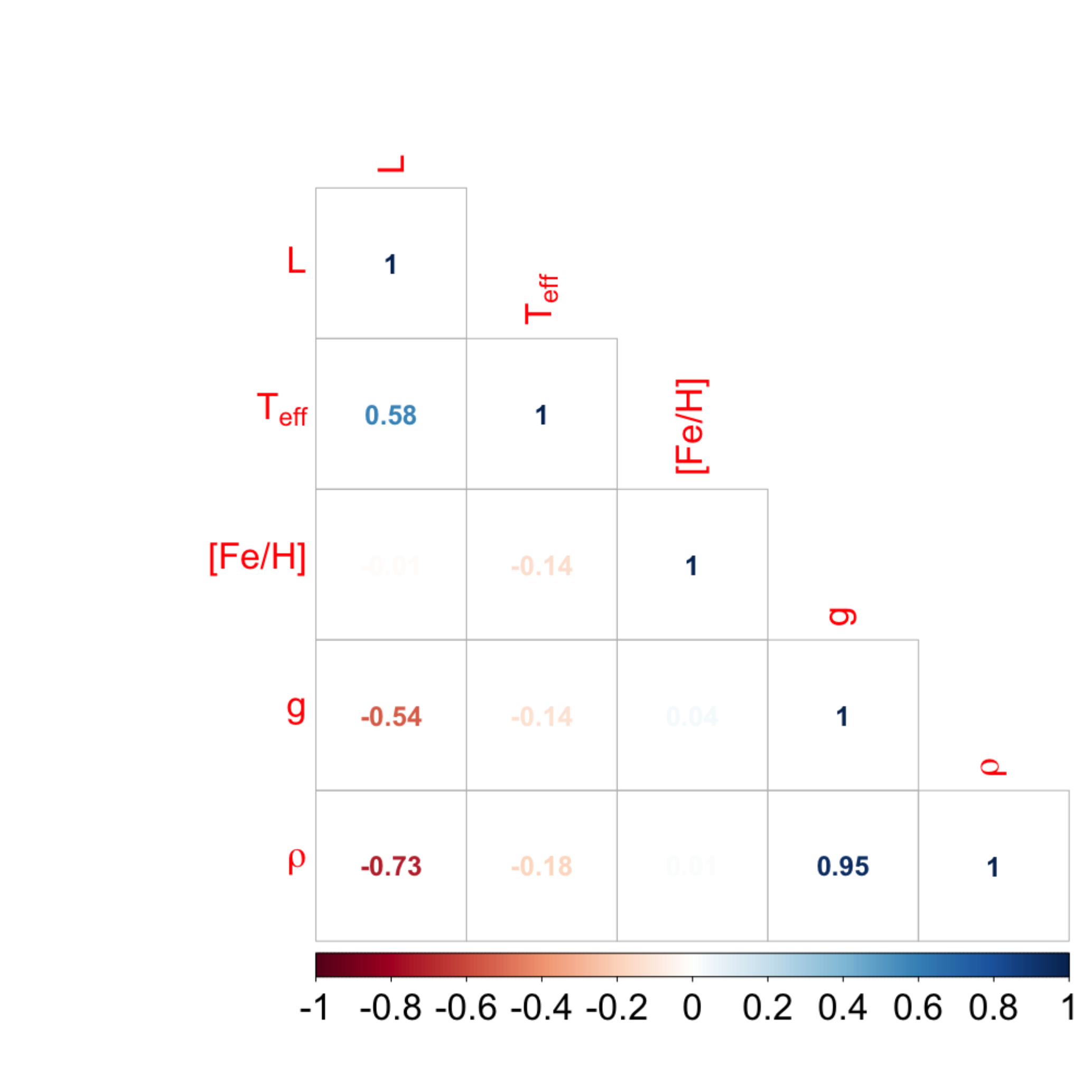}
 \caption{The Kendall-$\tau$ correlation coefficient of all the possible pairs of observables. Color grading from red to blue accounts for anti-correlation to direct correlation, respectively.}
 \label{fig:correl}
\end{figure}

There are combinations of variables that add little or no new information over a single variable. In Fig. \ref{fig:correl} we show the Kendall-$\tau$ correlation coefficient of all the possible pairs of observables. We find that there is only one strong correlation (larger than 0.75), i.e., there is only one obvious case of redundant variables. Gravity is highly correlated with density, as expected. Luminosity is also anti-correlated with density ($\tau$ =-0.73), close to our threshold. In Appendix \ref{sec:cross-cor} we show the scatter plots of these cross-correlations.
%
%

Therefore, we proceed to study all the variables as if they were independent except gravity-density. We have removed every relation where these two variables appear at the same time, since both provide redundant information. We have decided to keep those relations with luminosity and density at the same time since, although they are correlated, we estimate, looking at the scatter plot shown at the Appendix, that each one can provide some independent and complementary information.
%
%

We conclude this section by noting again that we are not focused on investigating physical clues from data, rather on obtaining relations that capture the source information provided by the methods described in section \ref{sec:data_sample}. When source data or information are available from those methods, we suggest to use them to estimate masses and radii. If such data are not available, the relations we present can offer similar but less precise estimations.

\subsection{Selection of the best subset}

For a given relation we select a subset of stars for the regression analysis that fulfill certain characteristics . The remaining stars are then used as the control group for studying the accuracy and precision of the relation.

The idea behind this selection is to balance the accuracy obtained when the variables with a better precision are used, with the precision obtained when the number of stars in the subsample is raised. We have found that a good balance between accuracy and precision in our results is reached when we select for the regression those stars with an uncertainty in $M$, $R$, $T_{\rm eff}$, ${\rm log}g$, and/or $\rho$ $\le 7\%$, and an uncertainty in $L$ $\le 10\%$. For example, if we are going to test the relation $M=f(T_{\rm eff},L)$, we first select the subset for the regression, which includes those stars fulfilling the requirements that $\Delta M$ and $\Delta T_{\rm eff}$ $\le 7\%$, and $\Delta L$ $\le 10\%$, leaving the rest of the stars as the control group. If the relation is $R=f(\rho)$, we select those stars fulfilling $\Delta R$ and $\Delta \rho$ $\le 7\%$, with the remaining stars again left as the controls.

This selection implies that the number of stars in the regression and control groups changes from one relation to another. At this point, we recall again that one of the features of this study is that we mix different techniques, trying to balance any possible bias of one technique with the unbiased determinations of the others. For every relation we present the percentage of stars characterized by the different techniques and with different spectral types (tables \ref{table:sesgo_method} and \ref{table:sesgo_type}). The percentages of the complete sample are displayed in Tables \ref{table:percentages_tech} and \ref{table:percentages_type} (see section \ref{sec:heter}).

\subsection{Analysis method}
\label{sec:analysis_method}

The use of an error-in-variables linear regression algorithm ensures a robust treatment of the measured uncertainties, and more reliable results compared with using only the central observed values, as is the case for the standard linear regressions.

Following \citet{Hansen12}, we use the error-in-variables model GLSME (Generalized Least Squares with Measurement Error):

\begin{eqnarray}
y &=& D\beta + r\nonumber \\
r &\backsim & N(0,V)\\
V &=& \sigma^2 T + V_e + {\rm Var}[U\beta|D] \nonumber
\label{eq:glsme}
\end{eqnarray}


\noindent where $y$ is a vector with the central values of the observed dependent variable, $D$ is a matrix with the central values of the observed independent variables, $\beta$ is a vector with the regression coefficients to be estimated, and $N(0,V)$ represents the normal distribution centered at zero having variance $V$. In the most general case, $V$ is comprised of the measurement uncertainties and the possible random effects of the model itself. $V_e$ is a matrix with the measurement errors of the dependent variable, $\sigma^2 T$ is a matrix with the residuals of the true dependent variable, that is, the impact of these possible random effects in the dependent variable. Finally, ${\rm Var}[U\beta|D]$ is a matrix counting for the independent variables uncertainties, and it contains $V_U$, the independent variables measurement errors, and $V_D$, the possible effects in the independent variables of a random term. In our case, we assume that, if there is a physical relation combining several variables, its application is deterministic. That is, there is not any additional random term. Therefore, $\sigma^2 T$ and $V_D$ =0, and only the measurement errors must be included in the study. Assuming that the published uncertainties of the different measurements correspond to $\sigma$ (unless they are explicitly informed), $V_e$ is an $n\times n$ diagonal matrix (with $n$ the number of stars used for obtaining the regression) with the $\sigma^2$ measurement uncertainties of the dependent variable. On the other hand, $V_U$ is a collection of $m$ $n\times n$ diagonal matrices (with $m$ the number of independent variables) with the $\sigma^2$ measurement uncertainties of the independent variables. For a more detailed analysis of the different components of the GLSME model, we refer the reader to the Appendix in \citet{Hansen12}

For every combination of variables (e.g. $M = f(T_{\rm eff},L)$), we construct all the possible alternatives including those with their logarithms (e.g. $M = f(T_{\rm eff},L)$, ${\rm log} M = f(T_{\rm eff},L)$, $M = f({\rm log} T_{\rm eff},L)$, ${\rm log} M = f({\rm log} T_{\rm eff},L)$, etc.). We then perform the error-in-variables linear regression, using de GLSME model, to obtain estimates of the regression coefficients $\beta$ and their uncertainties $\Delta\beta$. For each best-fitting relation we then extract the following summary statistics:

\begin{itemize}
\item The well-known $R^2$ statistic: This measures the percentage of the dependent variable variance explained by the linear regression, for the regression sample used to obtain the regression coefficients.

\item The Relative accuracy (Acc): For a given relation and control group (i.e., different from the regession sample used to obtain the linear relation), we have the expected values of the dependent variables ($y_{\rm fitted}$) and their ``real'' values ($\hat{y}$). We may therefore define the global relative accuracy of the linear regression as:
 \begin{equation}
 {\rm Acc} = {\rm Mean}\Big(\frac{|y_{i,{\rm fitted}} - \hat{y}_i|} 
 {\hat{y}_i}\Big) \times 100
 \label{eq:acc_def}
 \end{equation}

\item Relative precision (Prec): As per the above, we may also define the global relative precision of the linear regression as:
 \begin{equation}
 {\rm Prec} = {\rm Mean}\Big(\frac{\sigma_{i,{\rm fitted}}}{\hat{y}_i}\Big) \times 100
 \label{eq:prec_def}
 \end{equation}
where $\sigma_{i,{\rm fitted}}$ is the standard deviation when evaluating the relation for every element of the control group.
\end{itemize}

The standard deviation is obtained via error propagation. To estimate it for the relative precision of the dependent variable ($M$ or $R$) of the control group, only the central values of the independent variables are used. Therefore, the standard deviation ($\sigma_{i,{\rm fitted}}$) is a reflection only of the coefficient errors.


The selected combination for a given group of dependent and independent variables is that providing as high an $R^2$ and as low an Acc and Prec possible. Finally, only those relations with $R^2>0.85$ have been selected for further scrutiny.

\section{Relations found}
\label{sec:selection}

\begin{figure}
 \includegraphics[width=\columnwidth]{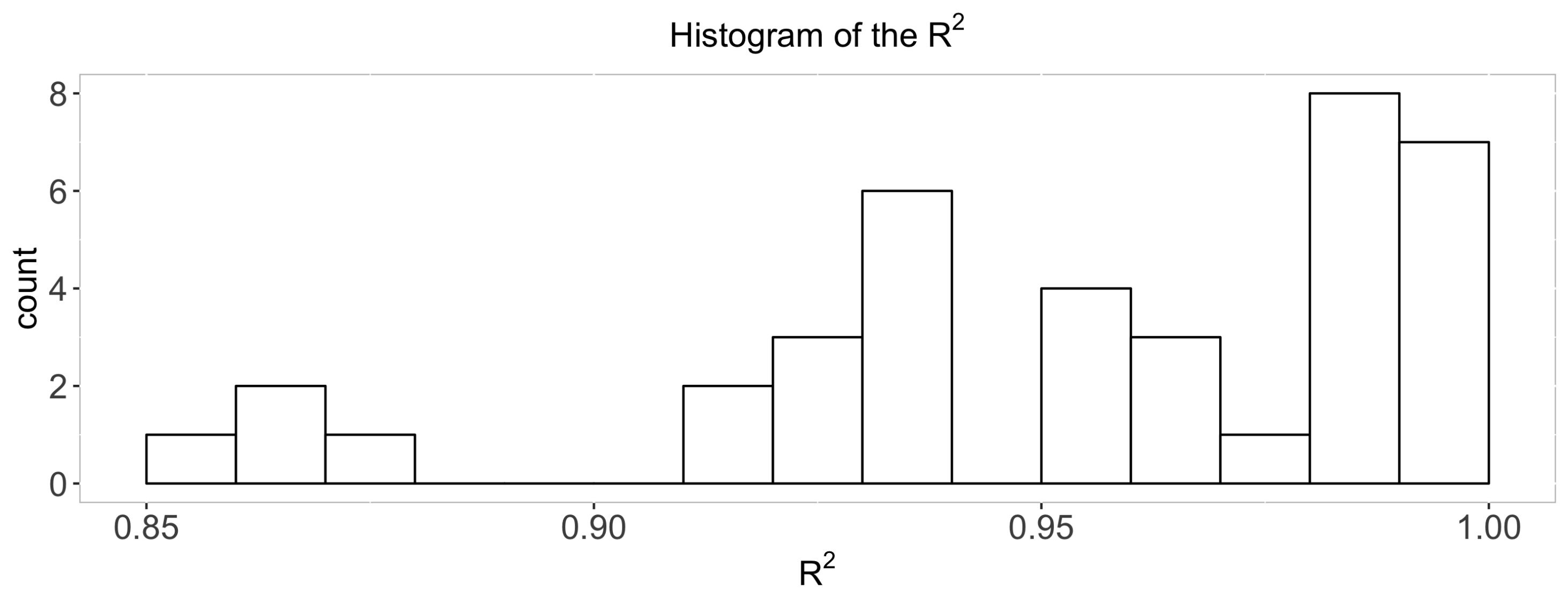}
 \caption{Histogram showing the $R^2$ of the relations selected.}
 \label{fig:R2}
\end{figure}

In Table \ref{table:result_r2_acc_prec} we present all the statistical characteristics of the selected relations. In terms of $R^2$, in Fig. \ref{fig:R2} we show a histogram of the values obtained. We see that most of the relations explain more than the 95$\%$ of the variance of the dependent variable, while 89$\%$ of them have a $R^2 > 0.9$. 

\startlongtable
\begin{deluxetable*}{ccccccc}
\tablecaption{Summary with the main statistics of the selected relations: $R^2$ is the percentage of variance in the dependent variable explained by the relation; Acc tot is the relative accuracy of the relation; Prec tot is the relative precision of the relation; Acc plan is the relative accuracy of the relation calculated with every star harboring planets in our sample; Prec plan is the relative precision of the relation calculated with every star harboring planets in our sample; Num. st control is the number of stars in the control group. \label{table:result_r2_acc_prec}
}
\tablehead{
\colhead{Rel} & \colhead{$R^2$} & \colhead{Acc. tot} & \colhead{Prec. tot} & \colhead{Acc. plan} & \colhead{Prec. plan} & \colhead{Num. st. control}} 
\startdata
  $M \backsim T_{\rm eff}$ & 0.87 & 8.29 & 0.44 & 7.45 & 0.49 & 125 \\ 
  ${\rm log}M \backsim {\rm log}L$ & 0.96 & 10.08 & 0.13 & 7.31 & 0.10 & 224 \\ 
  $R \backsim {\rm log}L$ & 0.86 & 13.55 & 0.08 & 8.29 & 0.07 & 122 \\ 
  ${\rm log}R \backsim {\rm log}g$ & 0.91 & 17.12 & 1.18 & 8.53 & 1.20 & 8 \\ 
  ${\rm log}R \backsim {\rm log}\rho$ & 0.98 & 2.86 & 0.23 & 2.61 & 0.19 & 81 \\ 
  $M \backsim T_{\rm eff} + {\rm [Fe/H]}$ & 0.86 & 6.23 & 1.29 & 6.09 & 1.37 & 180 \\ 
  ${\rm log}M \backsim {\rm log}L + {\rm [Fe/H]}$ & 0.94 & 9.91 & 0.88 & 6.66 & 0.94 & 200 \\ 
  ${\rm log}R \backsim {\rm log}L + {\rm [Fe/H]}$ & 0.93 & 9.06 & 0.17 & 5.73 & 0.14 & 161 \\ 
  ${\rm log}R \backsim {\rm log}g + {\rm [Fe/H]}$ & 0.96 & 4.96 & 1.95 & 4.50 & 1.98 & 138 \\ 
  ${\rm log}R \backsim {\rm log}\rho + {\rm [Fe/H]}$ & 0.99 & 2.87 & 0.27 & 2.48 & 0.23 & 174 \\ 
  ${\rm log}M \backsim T_{\rm eff} + {\rm log}L$ & 0.96 & 8.55 & 1.10 & 6.44 & 0.99 & 228 \\ 
  ${\rm log}R \backsim T_{\rm eff} + {\rm log}L$ & 0.99 & 5.43 & 0.89 & 2.30 & 0.76 & 126 \\ 
  $M \backsim T_{\rm eff} + {\rm log}g$ & 0.91 & 7.88 & 2.07 & 7.21 & 2.43 & 108 \\ 
  ${\rm log}R \backsim T_{\rm eff} + {\rm log}g$ & 0.99 & 7.61 & 1.47 & 3.16 & 1.48 & 8 \\ 
  $M \backsim T_{\rm eff} + {\rm log}\rho$ & 0.92 & 6.43 & 1.96 & 6.14 & 2.10 & 158 \\ 
  ${\rm log}R  \backsim T_{\rm eff} + {\rm log}\rho$ & 0.99 & 2.26 & 1.62 & 1.76 & 1.59 & 81 \\ 
  $M \backsim {\rm log}L + {\rm log}g$ & 0.93 & 8.92 & 3.29 & 6.46 & 3.98 & 203 \\ 
  ${\rm log}R \backsim {\rm log}L + {\rm log}g$ & 0.99 & 3.83 & 2.40 & 2.24 & 2.40 & 108 \\ 
  ${\rm log}M \backsim {\rm log}L + {\rm log}\rho$ & 0.96 & 7.76 & 0.70 & 4.67 & 0.56 & 163 \\ 
  ${\rm log}R \backsim {\rm log}L + {\rm log}\rho$ & 0.995 & 2.78 & 0.61 & 1.47 & 0.43 & 91 \\ 
  $M \backsim T_{\rm eff} + L + {\rm [Fe/H]}$ & 0.94 & 6.89 & 1.25 & 5.20 & 1.31 & 200 \\ 
  ${\rm log}R \backsim T_{\rm eff} + {\rm log}L + {\rm [Fe/H]}$ & 0.99 & 5.94 & 2.01 & 2.15 & 1.94 & 161 \\ 
  ${\rm log}M \backsim T_{\rm eff} + {\rm log}g + {\rm [Fe/H]}$ & 0.93 & 7.54 & 3.43 & 6.59 & 3.46 & 170 \\ 
  ${\rm log}R \backsim T_{\rm eff} + {\rm log}g + {\rm [Fe/H]}$ & 0.99 & 2.97 & 2.73 & 2.61 & 2.74 & 138 \\ 
  $M \backsim T_{\rm eff} + {\rm log}\rho + {\rm [Fe/H]}$ & 0.88 & 7.02 & 5.61 & 6.01 & 5.98 & 203 \\ 
  ${\rm log}R  \backsim T_{\rm eff} + {\rm log}\rho + {\rm [Fe/H]}$ & 0.997 & 1.69 & 2.67 & 1.35 & 2.61 & 174 \\ 
  ${\rm log}M \backsim {\rm log}L + g + {\rm [Fe/H]}$ & 0.97 & 8.32 & 0.83 & 4.79 & 0.84 & 190 \\ 
  ${\rm log}R \backsim {\rm log}L + g + {\rm [Fe/H]}$ & 0.96 & 5.54 & 0.74 & 3.72 & 0.73 & 158 \\ 
  $M \backsim {\rm log}L + \rho + {\rm [Fe/H]}$ & 0.93 & 7.82 & 1.72 & 3.93 & 1.82 & 204 \\ 
  ${\rm log}R \backsim {\rm log}L + {\rm log}\rho + {\rm [Fe/H]}$ & 0.997 & 2.21 & 0.85 & 1.20 & 0.60 & 176 \\  
  $M \backsim T_{\rm eff} + L + {\rm log}g$ & 0.94 & 8.65 & 2.46 & 8.38 & 2.95 & 203 \\ 
  ${\rm log}R \backsim T_{\rm eff} + L + {\rm log}g$ & 0.99 & 4.89 & 1.84 & 3.38 & 1.81 & 108 \\ 
  $M  \backsim T_{\rm eff} + {\rm log}L + \rho$ & 0.95 & 7.41 & 3.87 & 4.95 & 4.10 & 163 \\ 
  $R \backsim T_{\rm eff} + L + {\rm log}\rho$ & 0.98 & 3.19 & 1.99 & 2.61 & 2.39 & 91 \\ 
  $M \backsim T_{\rm eff} + L + {\rm log}g + {\rm [Fe/H]}$ & 0.95 & 8.49 & 4.45 & 6.62 & 4.97 & 190 \\ 
  ${\rm log}R \backsim T_{\rm eff} + {\rm log}L + g + {\rm [Fe/H]}$ & 0.99 & 5.59 & 2.79 & 1.67 & 2.67 & 158 \\ 
  $M  \backsim T_{\rm eff} + {\rm log}L + \rho + {\rm [Fe/H]}$ & 0.94 & 7.45 & 7.63 & 3.83 & 8.12 & 204 \\ 
  ${\rm log}R \backsim T_{\rm eff} + L + {\rm log}\rho + {\rm [Fe/H]}$ & 0.998 & 2.10 & 2.88 & 1.04 & 2.77 & 176 \\ 
\enddata
\end{deluxetable*}

If we take a look at the control groups, we can see that in most of the cases the number of stars in these groups is in the range [81, 228]. The statistical tests performed on these groups should as such, be reliable. There are two exceptions. The relations ${\rm log}R \backsim {\rm log}g$ and ${\rm log}R \backsim T_{\rm eff}+{\rm log}g$ have been tested with only 8 stars. Therefore, the Acc and Prec shown in these cases must be taken with caution.
%
%

\begin{figure}
 \includegraphics[width=\columnwidth]{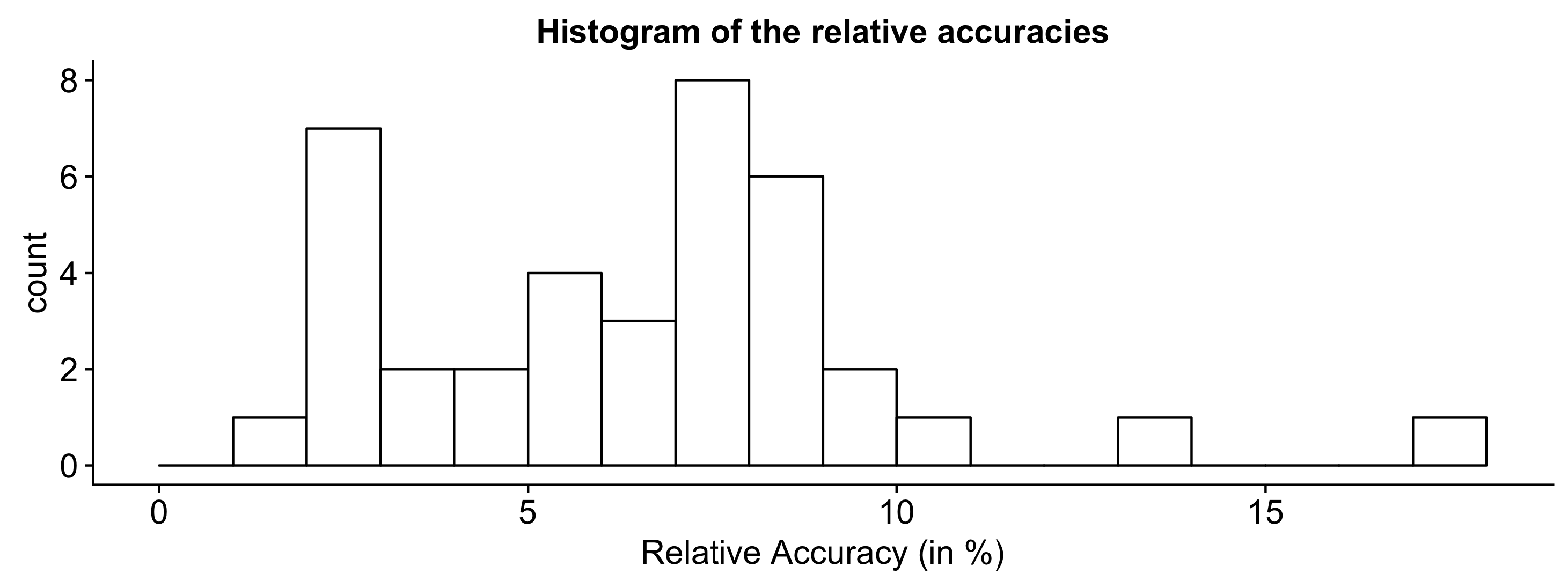}
 \caption{Histogram showing the Acc of the relations selected.}
 \label{fig:Acc}
\end{figure}

In Fig. \ref{fig:Acc} we show the histogram of the relative accuracies. All are lower than $10\%$ except for three cases: ${\rm log}M \backsim {\rm log}L$, $R \backsim {\rm log}L$ and ${\rm log}R \backsim {\rm log}g$. In general, the relative accuracy is lower (poorer) for relations using only one independent variable, as expected. Most of the relatives accuracies better than $5\%$ are related to the estimation of the radius. In general, the relations estimating the radius are more accurate than those estimating the mass (a mean value for all the relations of 5.3$\%$ ($R$) versus 7.98$\%$ ($M$)).

\begin{figure}
 \includegraphics[width=\columnwidth]{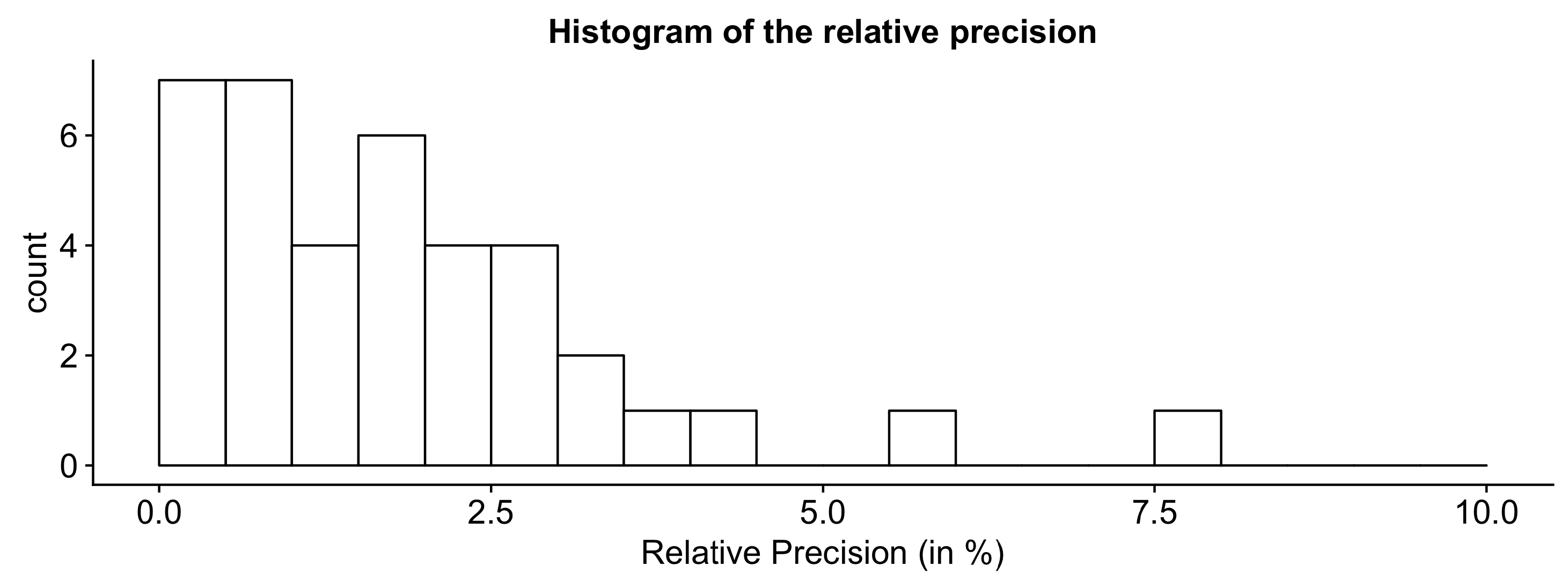}
 \caption{Histogram showing the Prec of the relations selected.}
 \label{fig:Prec}
\end{figure}

\begin{figure}
 \includegraphics[width=\columnwidth]{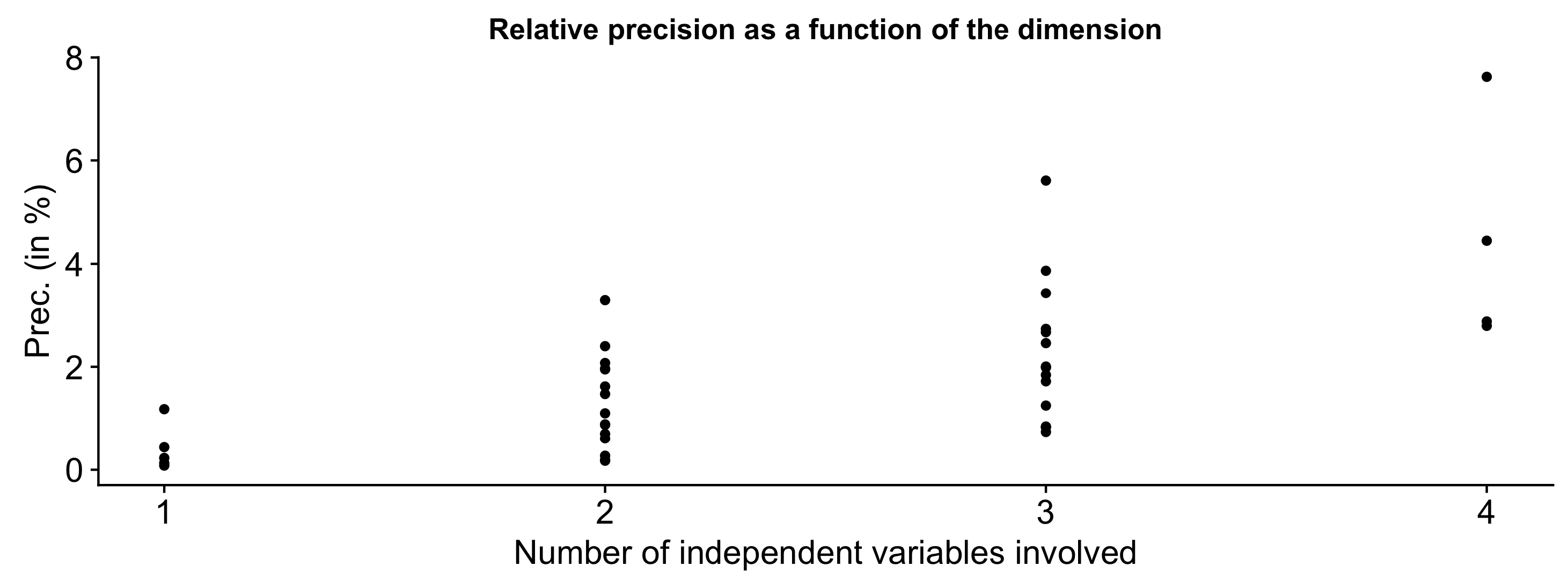}
 \caption{Relative precision as a function of the number of dimensions of the selected relations.}
 \label{fig:Prec_dim}
\end{figure}

In Fig. \ref{fig:Prec} we show the histogram of the relative precisions. Here we also find that most of the relations provide relative precisions better than $7.5\%$. In fact, 84$\%$ of the relations have a ${\rm Prec} < 3\%$. Note that these relative precisions take into account only the contribution of the errors in the regression coefficients. To obtain a realistic standard deviation for an estimation of a mass or radius we must add the uncertainty coming from the input variables. Therefore, the tight relative precisions shown in Fig.~\ref{fig:Prec} are good news. The relations for $R$ again provide better precisions than those for $M$ (a mean value for all the relations of 1.47$\%$ versus 2.33$\%$). Only one of the relations give relative precision worse than 7.5$\%$ (in particular 7.63$\%$): $M \backsim T_{\rm eff} + {\rm log}L + g + \rho + {\rm [Fe/H]}$, that is, a relation with a large number of dimensions. In fact, the relative precision in general deteriorates with the number of independent variables involved, as can be seen in Fig. \ref{fig:Prec_dim}. This precision deterioration is worst when the input uncertainties are included. The large the number of dimensions, the larger the impact of these uncertainties on the final uncertainty.
%
%
%

In table \ref{table:coefs} we show the best-fitting coefficients of the selected relations and their errors in the format $X(Y) \equiv X\times 10^{Y}$. The first column of the table describe the relation selected (e.g. $Z=f(X+Y)\equiv a\pm e_a+({\beta}_X\pm e_{{\beta}_X})X+({\beta}_Y\pm e_{{\beta}_Y})Y$). The coefficients shown are those multiplying the independent variables in the relations independently whether it is included as a logarithm or not.

In Table \ref{table:ranges} we show the ranges of validity of each relation. These ranges are set by the maximum and minimum values of each independent variable used in the relation (i.e., from the input data in the regression group used to obtain the relation). We see that, in general, the larger the number of independent variables involved, the narrower is the range of validity of the relation.

Finally, in the light of the high correlation found between gravity and density, we have obtained an error-in-variables regression model relating these two variables with the existing data sample. In this case, we have used the relation ${\rm logg} \backsim {\rm log}\rho$. A summary of the parameters of this relation can be found in Table \ref{table:g-rho}.

\begin{table}
\centering
\caption{Summary of the relation log g vs. log$\rho$. \label{table:g-rho}
}
\begin{tabular}{ccccccccc}
  \hline
Rel. & $R^2$ & \colhead{$a$} & \colhead{$e_a$} & \colhead{$\rho$} & \colhead{$e_\rho$} & Acc. & Prec. & \colhead{Num. st. control}\\ 
  \hline
${\rm logg} \backsim {\rm log}\rho$ & 0.99 & 4.4251 & 0.0008 & 0.5987 &  0.0013 & 2.66 & 0.28 & 74\\ 
   \hline
\end{tabular}
\end{table}

\section{Consistency checks}


\subsection{Ensuring the heterogeneity}
\label{sec:heter}

As noted previously, one of the features of this work is that we have used a heterogeneous data, in terms of the techniques used, since this can in principle reduce the influence of possible biases inherent in the observations, and reduction and analysis methods. As described in section \ref{sec:selection}, to extract the different relations we use a subset of stars fulfilling certain criteria. Here, we test whether these selections affect the heterogeneity of each regression sample.

\startlongtable
\begin{deluxetable*}{cccc}
\tablecaption{Percentage of stars in the different subsets used for obtaining the regressions characterized by different techniques. \label{table:sesgo_method}
}
\tablehead{
\colhead{Rel} & \colhead{A} & \colhead{EB} & \colhead{I}}
\startdata
$M \backsim T_{\rm eff}$ &  57.4 & 40.9 & 1.7 \\ 
  ${\rm log}M \backsim {\rm log}L$ &  66.7 & 31.3 & 2.0 \\ 
  $R \backsim {\rm log}L$ &  71.4 & 27.0 & 1.6 \\ 
  ${\rm log}R \backsim {\rm log}g$ &  65.2 & 34.8 &   -- \\ 
  ${\rm log}R \backsim {\rm log}\rho$ &  99.5 & 0.5 &   -- \\ 
  $M \backsim T_{\rm eff}+{\rm [Fe/H]}$ &  82.8 & 14.3 & 2.9 \\ 
  ${\rm log}M \backsim {\rm log}L+{\rm [Fe/H]}$ &  86.3 & 10.6 & 3.1 \\ 
  ${\rm log}R \backsim {\rm log}L+{\rm [Fe/H]}$ &  87.8 & 9.4 & 2.8 \\ 
  ${\rm log}R \backsim {\rm log}g+{\rm [Fe/H]}$ &  86.7 & 13.3 &   -- \\ 
  ${\rm log}R \backsim {\rm log}\rho+{\rm [Fe/H]}$ & 100.0 &  -- &   -- \\ 
  ${\rm log}M \backsim T_{\rm eff}+{\rm log}L$ &  67.2 & 30.8 & 2.0 \\ 
  ${\rm log}R \backsim T_{\rm eff}+{\rm log}L$ &  71.8 & 26.5 & 1.7 \\ 
  $M \backsim T_{\rm eff}+{\rm log}g$ &  60.0 & 40.0 &   -- \\ 
  ${\rm log}R \backsim T_{\rm eff}+{\rm log}g$ &  65.2 & 34.8 &   -- \\ 
  $M \backsim T_{\rm eff}+{\rm log}\rho$ &  99.3 &  0.7 &   -- \\ 
  ${\rm log}R  \backsim T_{\rm eff}+{\rm log}\rho$ &  99.5 &  0.5 &   -- \\ 
  $M \backsim {\rm log}L+{\rm log}g$ &  70.4 & 29.6 &   -- \\ 
  ${\rm log}R \backsim {\rm log}L+{\rm log}g$ &  74.4 & 25.6 &   -- \\ 
  ${\rm log}M \backsim {\rm log}L+{\rm log}\rho$ &  99.3 &  0.7 &   -- \\ 
  ${\rm log}R \backsim {\rm log}L+{\rm log}\rho$ &  99.4 &  0.6 &   -- \\ 
  $M \backsim T_{\rm eff}+L+{\rm [Fe/H]}$ &  86.3 & 10.6 & 3.1 \\ 
  ${\rm log}R \backsim T_{\rm eff}+{\rm log}L+{\rm [Fe/H]}$ &  87.8 & 9.4 & 2.8 \\ 
  ${\rm log}M \backsim T_{\rm eff}+{\rm log}g+{\rm [Fe/H]}$ &  85.4 & 14.6 &   -- \\ 
  ${\rm log}R \backsim T_{\rm eff}+{\rm log}g+{\rm [Fe/H]}$ &  86.7 & 13.3 &   -- \\ 
  $M \backsim T_{\rm eff}+{\rm log}\rho+{\rm [Fe/H]}$ & 100.0 &    -- &   -- \\ 
  ${\rm log}R  \backsim T_{\rm eff}+{\rm log}\rho+{\rm [Fe/H]}$ & 100.0 & -- &   -- \\ 
  ${\rm log}M \backsim {\rm log}L+g+{\rm [Fe/H]}$ &  89.3 & 10.7 &   -- \\ 
  ${\rm log}R \backsim {\rm log}L+g+{\rm [Fe/H]}$ &  90.3 & 9.7 &   -- \\ 
  $M \backsim {\rm log}L+\rho+{\rm [Fe/H]}$ & 100.0 &    -- &   -- \\ 
  ${\rm log}R \backsim {\rm log}L+{\rm log}\rho+{\rm [Fe/H]}$ & 100.0 &  -- &   -- \\ 
  $M \backsim T_{\rm eff}+L+{\rm log}g$ &  70.4 & 29.6 &   -- \\ 
  ${\rm log}R \backsim T_{\rm eff}+L+{\rm log}g$ &  74.4 & 25.6 &   -- \\ 
  $M  \backsim T_{\rm eff}+{\rm log}L+\rho$ &  99.3 &  0.7 &   -- \\ 
  $R \backsim T_{\rm eff}+L+{\rm log}\rho$ &  99.4 &  0.6 &   -- \\ 
  $M \backsim T_{\rm eff}+L+{\rm log}g+{\rm [Fe/H]}$ &  89.3 &  10.7 &   -- \\ 
  ${\rm log}R \backsim T_{\rm eff}+{\rm log}L+g+{\rm [Fe/H]}$ &  90.3 &  9.7 &   -- \\
  $M \backsim T_{\rm eff}+{\rm log}L+\rho+{\rm [Fe/H]}$ &  100.0 &  -- &   -- \\ 
  ${\rm log}R \backsim T_{\rm eff}+{\rm log}L+{\rm log}\rho+{\rm [Fe/H]}$ &  100.0 &  -- &   -- \\ 
\enddata
\end{deluxetable*}

\startlongtable
\begin{deluxetable*}{cccccc}
\tablecaption{Percentage of stars in the different subsets used for obtaining the regressions with different spectral types. \label{table:sesgo_type}}
\tablehead{
\colhead{Rel} & \colhead{B} & \colhead{A} & \colhead{F} & \colhead{G} & \colhead{K}}
\startdata
  $M \backsim  T_{\rm eff}$ & 2.3 & 13.2 & 60.2 & 23.0 & 1.3 \\ 
  ${\rm log}M \backsim  {\rm log}L$ & 0.8 & 9.5 & 64.9 & 23.4 & 1.4 \\ 
  $R \backsim  {\rm log}L$ & 0.7 & 8.4 & 66.3 & 23.4 & 1.2 \\ 
  ${\rm log}R \backsim  {\rm log}g$ & 1.4 & 11.7 & 63.0 & 22.8 & 1.1 \\ 
  ${\rm log}R \backsim  {\rm log}\rho$ &    -- & 0.5 & 72.2 & 27.0 & 0.3 \\ 
  $M \backsim  T_{\rm eff}+{\rm [Fe/H]}$ &  0.3 & 1.5 & 67.6 & 28.9 & 1.7 \\ 
  ${\rm log}M \backsim  {\rm log}L+{\rm [Fe/H]}$ &    -- & 1.2 & 68.6 & 28.3 & 1.9 \\ 
  ${\rm log}R \backsim  {\rm log}L+{\rm [Fe/H]}$ &    -- & 1.1 & 68.1 & 29.1 & 1.7 \\ 
  ${\rm log}R \backsim  {\rm log}g+{\rm [Fe/H]}$ &  0.3 & 1.4 & 68.3 & 28.6 & 1.4 \\ 
  ${\rm log}R \backsim  {\rm log}\rho+{\rm [Fe/H]}$ &    -- &   -- & 69.5 & 30.1 & 0.4 \\ 
  ${\rm log}M \backsim  T_{\rm eff}+{\rm log}L$ & 0.8 & 9.4 & 64.8 & 23.6 & 1.4 \\ 
  ${\rm log}R \backsim  T_{\rm eff}+{\rm log}L$ & 0.7 & 8.2 & 66.3 & 23.6 & 1.2 \\ 
  $M \backsim  T_{\rm eff}+{\rm log}g$ & 1.1 & 13.3 & 61.4 & 22.9 & 1.3 \\ 
  ${\rm log}R \backsim  T_{\rm eff}+{\rm log}g$ & 1.4 & 11.7 & 63.1 & 22.7 & 1.1 \\ 
  $M \backsim  T_{\rm eff}+{\rm log}\rho$ &    -- & 0.7 & 69.7 & 29.3 & 0.3 \\ 
  ${\rm log}R  \backsim  T_{\rm eff}+{\rm log}\rho$ &    -- & 0.5 & 72.2 & 27.0 & 0.3 \\ 
  $M \backsim  {\rm log}L+{\rm log}g$ & 0.2 & 9.3 & 65.7 & 23.5 & 1.3 \\ 
  ${\rm log}R \backsim  {\rm log}L+{\rm log}g$ & 0.2 & 8.2 & 67.5 & 23.0 & 1.1 \\ 
  ${\rm log}M \backsim  {\rm log}L+{\rm log}\rho$ &    -- & 0.7 & 69.5 & 29.4 & 0.4 \\ 
  ${\rm log}R \backsim  {\rm log}L+{\rm log}\rho$ &    -- & 0.6 & 72.0 & 27.1 & 0.3 \\ 
  $M \backsim  T_{\rm eff}+L+{\rm [Fe/H]}$ &    -- & 1.2 & 68.6 & 28.3 & 1.9 \\ 
  ${\rm log}R \backsim  T_{\rm eff}+{\rm log}L+{\rm [Fe/H]}$ &    -- & 1.1 & 68.1 & 29.1 & 1.7 \\ 
  ${\rm log}M \backsim  T_{\rm eff}+{\rm log}g+{\rm [Fe/H]}$ &  0.3 & 1.6 & 68.2 & 28.3 & 1.6 \\ 
  ${\rm log}R \backsim  T_{\rm eff}+{\rm log}g+{\rm [Fe/H]}$ &  0.3 & 1.4 & 68.3 & 28.6 & 1.4 \\ 
  $M \backsim  T_{\rm eff}+{\rm log}\rho+{\rm [Fe/H]}$ &    -- &   -- & 68.4 & 31.2 & 0.4 \\ 
  ${\rm log}R  \backsim  T_{\rm eff}+{\rm log}\rho+{\rm [Fe/H]}$ &    -- &   -- & 69.5 & 30.1 & 0.4 \\ 
  ${\rm log}M \backsim  {\rm log}L+g+{\rm [Fe/H]}$ &    -- & 1.3 & 69.1 & 27.9 & 1.7 \\ 
  ${\rm log}R \backsim  {\rm log}L+g+{\rm [Fe/H]}$ &    -- & 1.2 & 69.1 & 28.2 & 1.5 \\ 
  $M \backsim  {\rm log}L+\rho+{\rm [Fe/H]}$ &    -- &   -- & 68.6 & 31.0 & 0.4 \\ 
  ${\rm log}R \backsim  {\rm log}L+{\rm log}\rho+{\rm [Fe/H]}$ &    -- &   -- & 69.6 & 30.0 & 0.4 \\ 
  $M \backsim  T_{\rm eff}+L+{\rm log}g$ & 0.2 & 9.3 & 65.7 & 23.5 & 1.3 \\ 
  ${\rm log}R \backsim  T_{\rm eff}+L+{\rm log}g$ & 0.2 & 8.2 & 67.5 & 23.0 & 1.1 \\ 
  $M  \backsim  T_{\rm eff}+{\rm log}L+\rho$ &    -- & 0.7 & 69.5 & 29.4 & 0.4 \\ 
  $R \backsim  T_{\rm eff}+L+{\rm log}\rho$ &    -- & 0.6 & 72.0 & 27.1 & 0.3 \\   
  $M \backsim  T_{\rm eff}+L+{\rm log}g+{\rm [Fe/H]}$ &    -- & 1.3 & 69.1 & 27.9 & 1.7 \\ 
  ${\rm log}R \backsim  T_{\rm eff}+{\rm log}L+g+{\rm [Fe/H]}$ &    -- & 1.2 & 69.1 & 28.2 & 1.5 \\ 
  $M  \backsim  T_{\rm eff}+{\rm log}L+\rho+{\rm [Fe/H]}$ &    -- &   -- & 68.6 & 31.0 & 0.4 \\ 
  ${\rm log}R \backsim  T_{\rm eff}+L+{\rm log}\rho+{\rm [Fe/H]}$ &    -- &   -- & 69.6 & 30.0 & 0.4 \\ 
\enddata
\end{deluxetable*}

In table \ref{table:sesgo_method} we display the percentage of stars from asteroseismology (A), eclipsing binaries (EB) and interferometry (I) in the regression sample used to obtain each relation. We see that there are two groups of relations: those with a balance of techniques similar to that of the complete sample (see Table \ref{table:percentages_tech}) and those where most of the stars (or the 100$\%$) come from the asteroseismic subsample. The reason for this difference is the presence or absence of $\rho$ as an independent variable. Asteroseismology provides a strong constraint on density directly from observations. Therefore, those relations including the density {\it may} be impacted by any possible bias coming from this technique. The rest of the relations are well balanced. The number of stars coming from interferometry is small, and the presence of them in the subsample does not have a significant impact on the statistical balance.

We have also looked carefully at the impact of the stellar spectral type. In Table \ref{table:sesgo_type} we present the percentage of stars of different spectral type that feature in the regression samples for each relation. We see that the main contribution comes from F-stars, followed by G-stars (with percentages similar to the global sample; see Table \ref{table:percentages_type}). The rest of the spectral types have smaller contributions depending on the relation studied, but the balance and the contribution of different spectral types is generally similar throughout. That said, we note two small biases: (i) cool stars (K stars and the only M star of the sample) have in general a small presence in the subsamples; and (ii) when the density is in the relation, there is a larger contribution of F and G-stars, since asteroseismology provides most of its data for these stellar types. 

\subsection{Linear regressions consistency}
\label{sec:reg_cons}

In addition to using $R^2$, relative accuracy and relative precision as main statistics for studying the quality of the regressions, we have also developed additional consistency tests to ensure that the linear regressions are representative of the observational data.

\begin{figure}
 \includegraphics[width=\columnwidth]{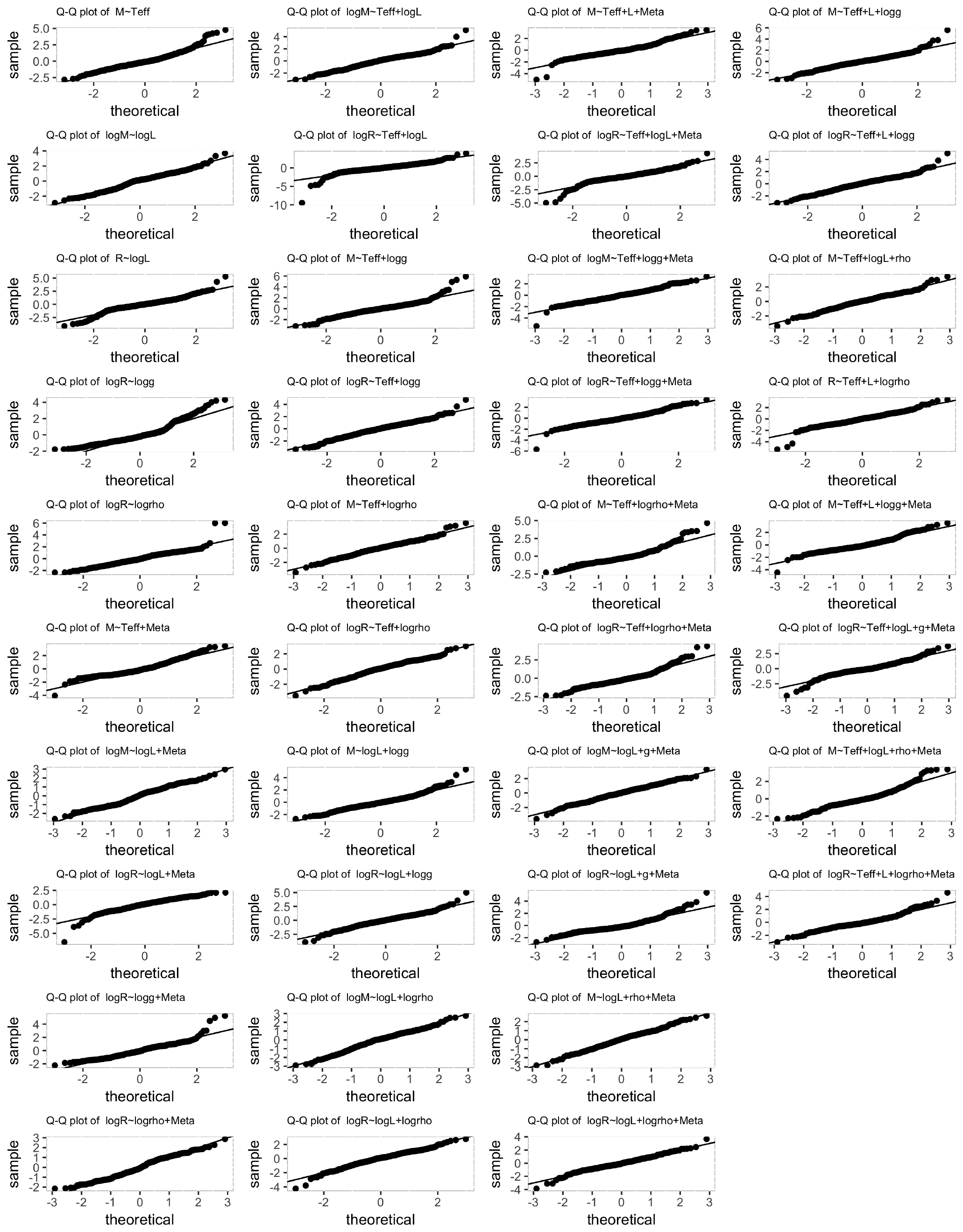}
 \caption{q-q plots of the 38 relations. See text for details.}
 \label{fig:Q-Q}
\end{figure}

In Fig. \ref{fig:Q-Q} we show q-q plots of the 38 selected relations selected in a form to show whether the standardized residuals are normally distributed. The ordered standardized residuals are plotted on the ordinate of each plot, while the expected order statistics from a standard normal distribution are on the abscissa. Points close to the straight line are consistent with a normal distribution. In this figure we can see that all the relations do in general follow this straight line. Therefore, the use of linear regression is justified. Only in a few cases are there extreme departures from this straight line, but, in general, the departures are not statistically significant.

\begin{figure}
 \includegraphics[width=\columnwidth]{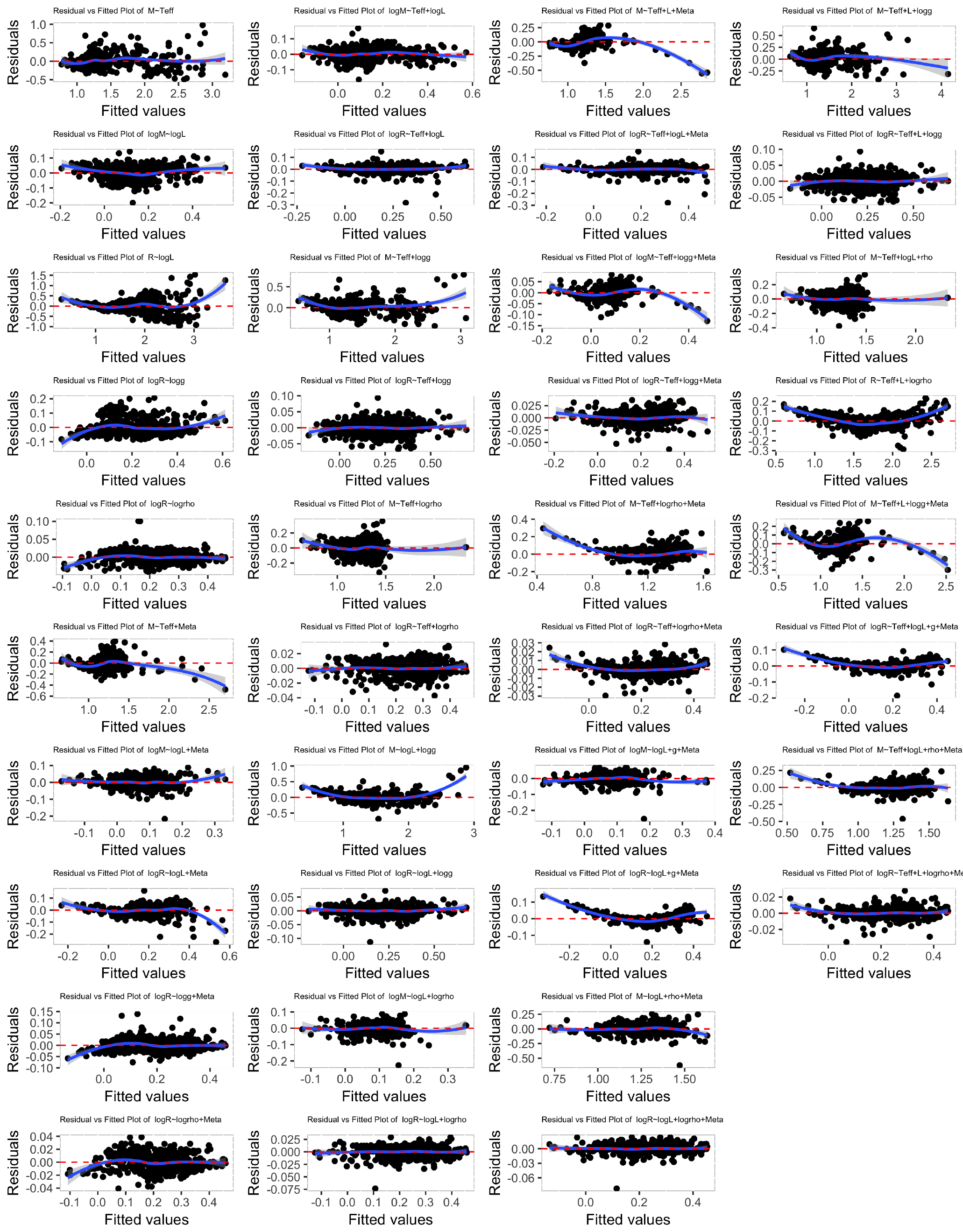}
 \caption{Residuals vs. Fitted values of the 38 relations. The line is a LOESS curve to guide the eye. See text for details.}
 \label{fig:Residuals}
\end{figure}

In Fig. \ref{fig:Residuals} we present the residuals as a function of the fitted values of these 38 relations. Any clear trend in these residuals can be a signature of inaccurate or inefficient regression. We have also added a LOESS (local polynomial regression) curve to guide the eye. In this figure we can see that, in general, there are no clear trends in the distributions of the residuals. In every plot, the main body of points is randomly distributed around the value zero. There are a few extreme cases, but they contain only a small percentage of the observational set. The impact of these values on the regression coefficients is analyzed in the next figure. 

\begin{figure}
 \includegraphics[width=\columnwidth]{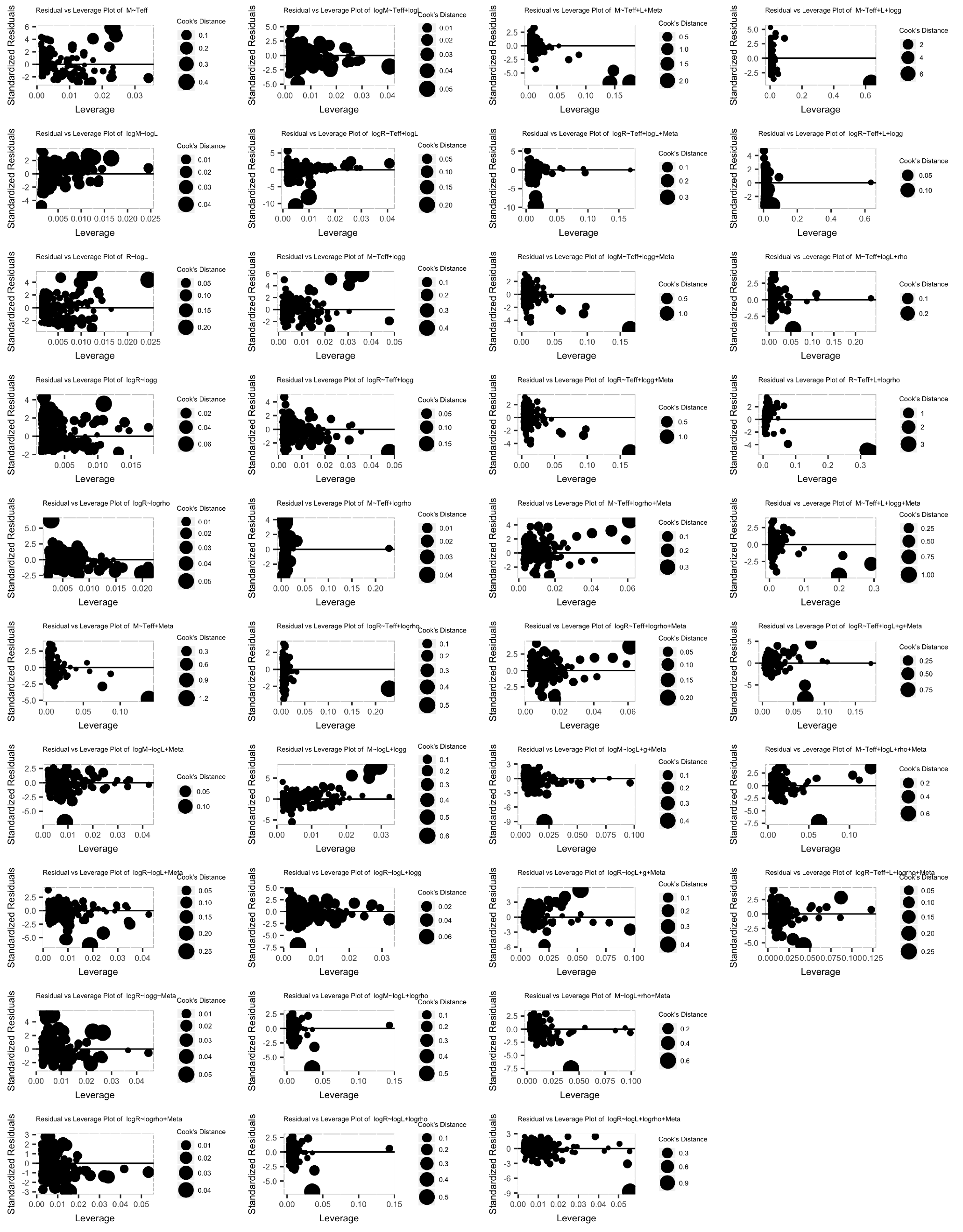}
 \caption{Residuals vs. Leverage plot of the 38 relations. The size of the points is proportional to the Cook's distance. See text for details.}
 \label{fig:Cook}
\end{figure}

In Fig. \ref{fig:Cook} we represent a final and more complex consistency test. Here we analyze the influence of every observational point in the regression coefficients. This influence is calculated using the Cook's distance \citep[D$_i$]{Reg_R}. This distance is calculated as a combination of the residual and the leverage (or how isolated a value is) for every point. The plots of Fig. \ref{fig:Cook} show the standardized residuals as a function of the leverage, and the Cook's distance is represented by the size of the points. According to \citet{Weisberg05}, "... if the largest value of D$_i$ is substantially less than one, deletion of a case will not change the estimate ... by much". Following this interpretation, only in four cases we have some points with D$_i>1$, and another two with some points with D$_i$ close to 1. In all cases, these points have large leverages, that is, they have a large influence on the estimates because they are extreme points isolated from the rest. This means that in these cases there are zones in this parameter space poorly sampled by our set, pointing where we must focus on improving our sampling.

\subsection{Influence of the definition of the Post-MS}
\label{sec:MS}

In section \ref{sec:data_sample}, when we described the data sampling, we mentioned the number of stars labeled as Main-Sequence (MS). There we explained that we used the evolutionary tracks with solar metallicity described in \citet{Rodrigues17} for this classification. The observational classification of a star as MS or post-MS is not a trivial task. Therefore, we have analyzed the impact on our results of using different tracks and physics to make this selection.

We have used tracks described in \citet{Rodrigues17} using the same physics, but with different metallicities in a range $Z=[0.00176, 0.0553]$. In addition, we have also used tracks that include diffusion and also cover a wider range of metallicities, ranging in $Z=[0.00002, 0.06215]$. In every case, the free parameters were calibrated so that a $1M_{\odot}$ model describes the Sun at solar age.

For each track, we select the position in the $T_{\rm eff} - {\rm log}g$ diagram where the star leaves the MS. The spread given by the different adopted model grids enables us to construct a probability distribution for the classification.

Using a Monte Carlo method, we have constructed up to 100 possible classifications of our 934 stars, resulting in 100 different subsets of stars classified as MS, and tested the impact of these different possible classifications on our results. Here, we show the impact for one of the relations of Table \ref{table:result_r2_acc_prec}: $M = a + b*T_{\rm eff} + c*L +  d*[Fe/H]$. The results obtained are shown in Table \ref{table:influence_ms}. Here we see the values obtained for the coefficients, their errors, and the statistics used for characterizing the goodness of the fit. "Mean" is the mean of each element over the 100 realizations; "S.D." is the standard deviation of these 100 realizations; and "Real" is the value we have obtained with our reference classification.

\begin{table}
\centering
\caption{Summary of the results obtained with the different classifications of the stars in our sampling as MS or post-MS. See text for details. \label{table:influence_ms}
}
\begin{tabular}{cccc}
  \hline
  Element    &  Mean        &   S.D.   &   Real\\
  \hline

$a$    &         -3.47(-1)&  6(-3)  &  -3.16(-1)\\
$T_{\rm eff}$  &       2.353(-4) &  1.2(-6)&   2.289(-4)\\
$L$   &          3.61(-2)&  5(-4) &   3.88(-2)\\
$[Fe/H]$      &       1.50(-1) &  3(-4) &   1.31(-1)\\
$e_a$ &          8.77(-3) &  7(-5)&    1.0(-2)\\
$e_{T_{\rm eff}}$  &      1.526(-6) &  1.5(-8) &   1.8(-6)\\
$e_{L}$     &      4.03(-4)&   1.0(-5) &   5(-4)\\
$e_{[Fe/H]}$       &    3.893(-3) &  3(-7) &   4(-3)\\
$R^2$       &     0.958 &   0.003&   0.94\\
Acc. tot &  7.2&   0.3 &   6.89\\
Prec. tot & 4.76&   0.23 &   2.00\\
Acc. plan &   6.08 &  0.16 &   5.20\\
Prec. plan &  4.81&   0.23 &   2.03\\
\hline
\end{tabular}
\end{table}

It is evident that the impact on the results of changes to the classification is small.

\subsection{Results obtained using other methods}

We have compared our results with those coming from the use of the standard linear regression (SR), and from a Random Forest model.

\begin{figure}
 \includegraphics[width=\columnwidth]{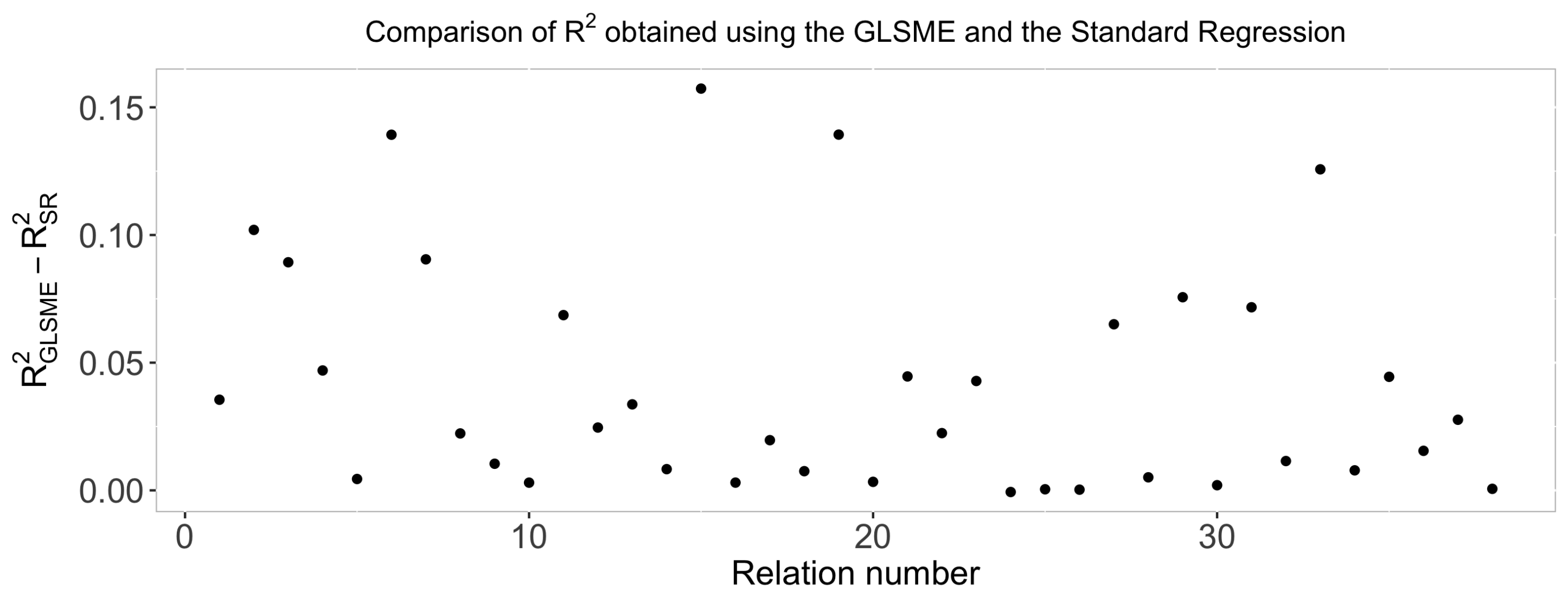}
 \caption{Differences between the statistic $R^2$ obtained with the GLSME and the standard regression algorithms for the 38 selected relations.}
 \label{fig:lm_glsme_r2}
\end{figure}

\begin{figure}
 \includegraphics[width=\columnwidth]{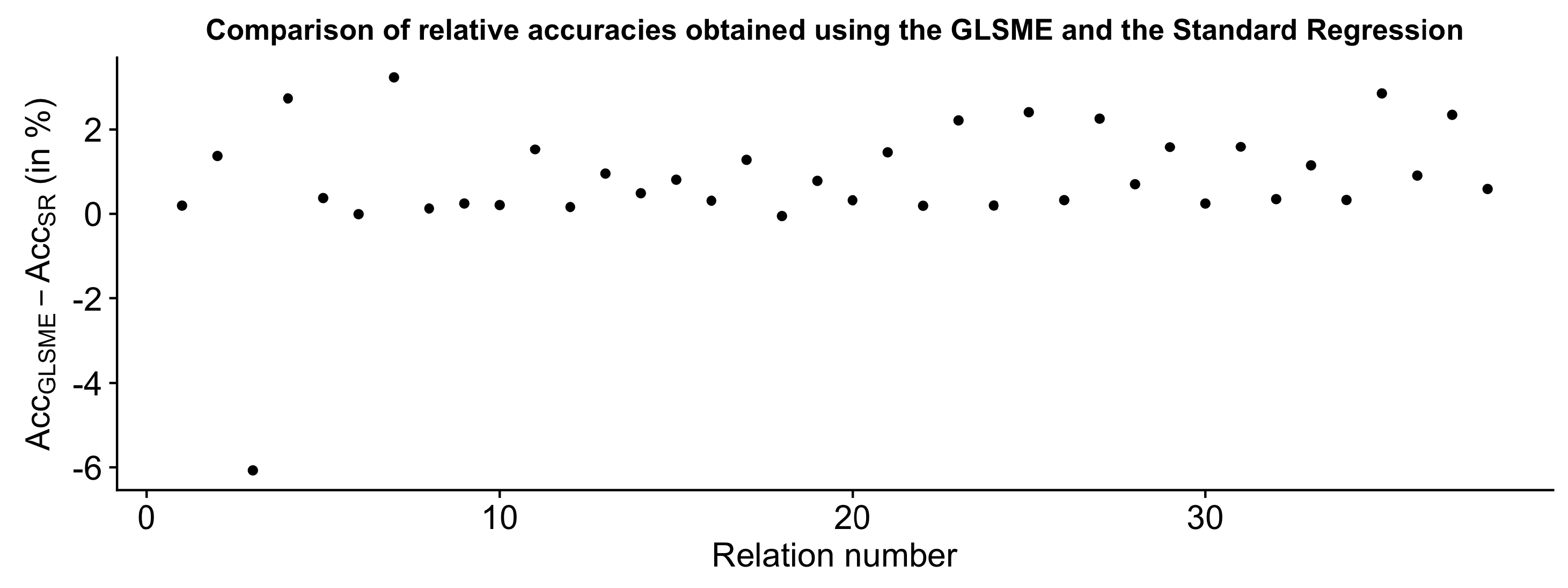}
 \caption{Differences between the relative accuracies obtained with the GLSME and the standard regression algorithms for the 38 selected relations.}
 \label{fig:lm_glsme_acc}
\end{figure}

\begin{figure}
 \includegraphics[width=\columnwidth]{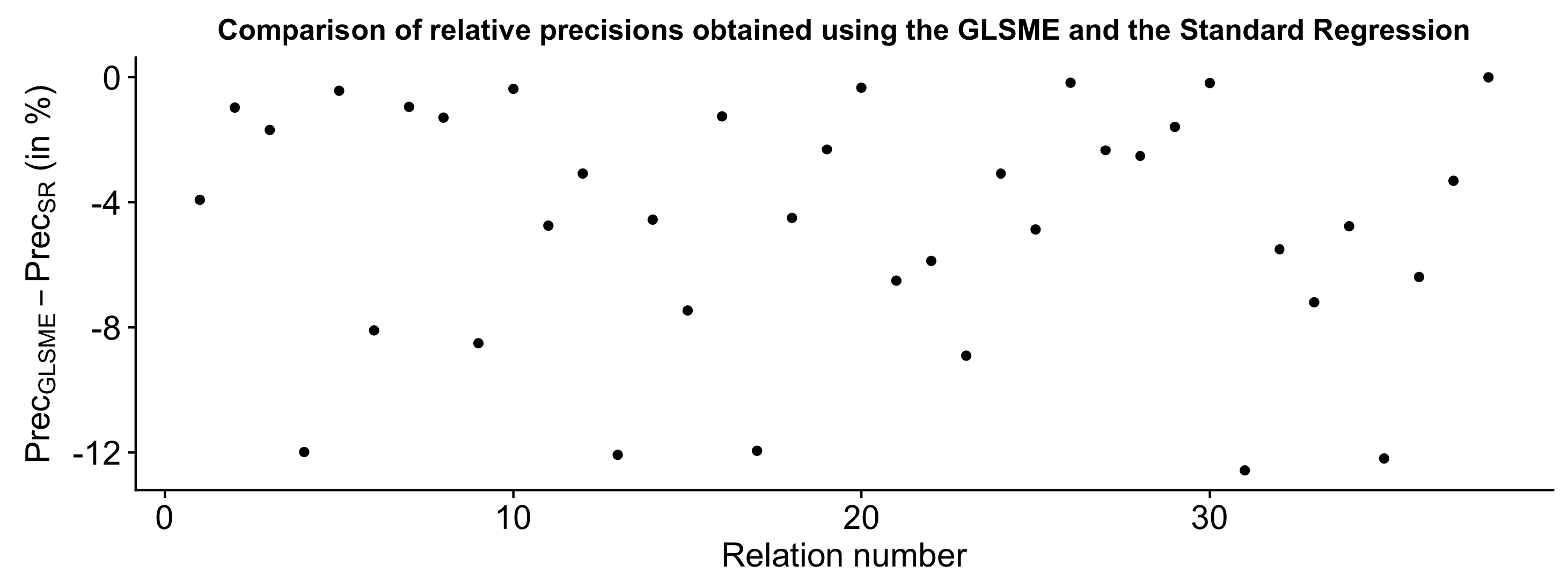}
 \caption{Differences between the relative precisions obtained with the GLSME and the standard regression algorithms for the 38 selected relations.}
 \label{fig:lm_glsme_prec}
\end{figure}

The most common algorithm for fitting a model to a group of data is the standard linear regression. We have repeated our analyses using standard linear regressions for the 38 selected relations. The comparisons are displayed in Figs. \ref{fig:lm_glsme_r2} to \ref{fig:lm_glsme_prec}. In all cases, a value >0 means that GLSME results are larger than SR ones (respectively <0 and lower values). In Fig. \ref{fig:lm_glsme_r2} we show the difference between the $R^2$ obtained with the GLSME algorithm (see Table \ref{table:result_r2_acc_prec}) and the $R^2$ obtained with standard linear regression (denoted here by $R_{\rm SR}^2$). The differences are small, with a mean offset of 0.04 and a maximum value of 0.157. Therefore, both algorithms provide models explaining almost the same dependent variable variance with almost all $R^2 > R^2_{\rm SR}$, that is, GLSME explains more variance of the dependent variable than the Standard Regression. In Fig. \ref{fig:lm_glsme_acc} we compare the relative accuracies coming from both algorithms. The differences are again small, with a mean difference of 0.80$\%$ and a maximum difference of 3.23$\%$, with an outlier of -6.07$\%$ on the relation $R ~ {\rm log}L$. Therefore, both algorithms provide similar relative accuracies, especially when describing the radius. Finally, in Fig. \ref{fig:lm_glsme_prec} we compare the relative precisions. Here we find the largest differences, always in favor of the GLSME results, as expected. No clear trends can be identified at this Fig. The mean difference in precision is of -4.64$\%$. This is critical specially for the relations with the larger number of dimensions, since the inclusion of the input uncertainties deteriorates even more the final precision.


\begin{figure*}
\includegraphics[width=\columnwidth]{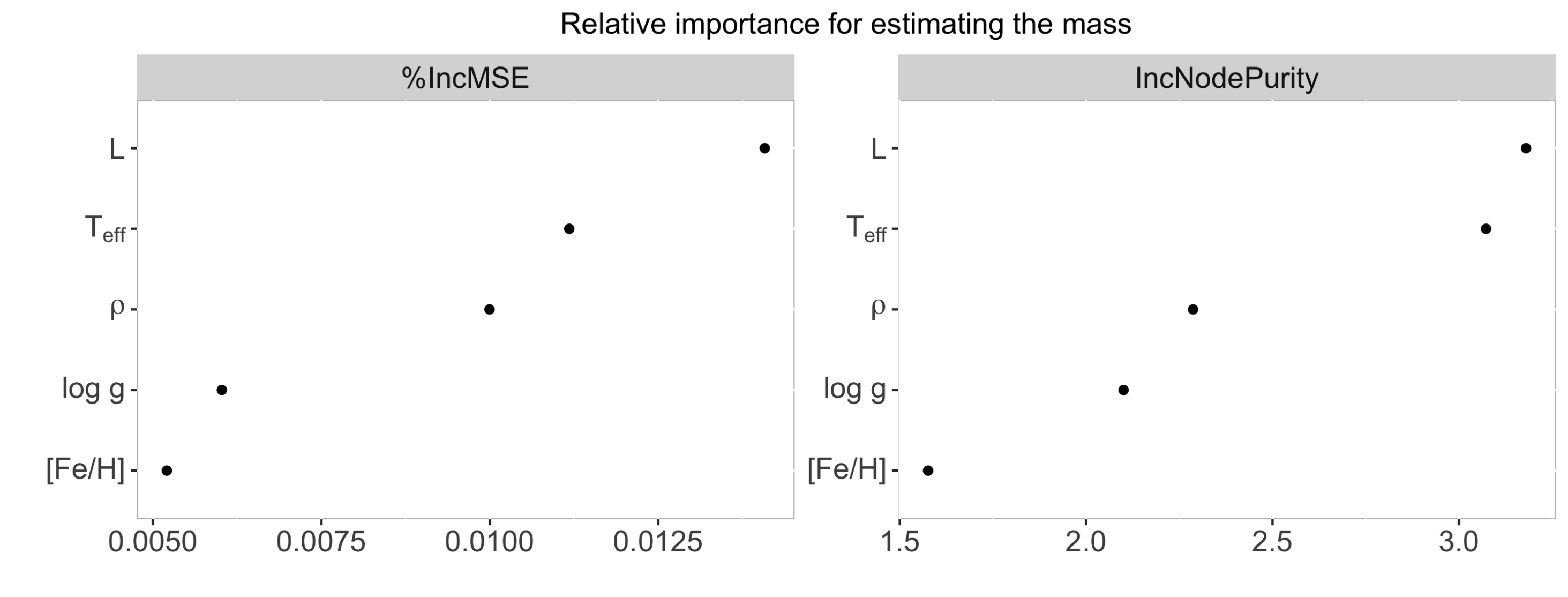}
\caption{Relative importance of the different independent variables in the mass estimations obtained using the Random Forest model. See text for details.}
 \label{fig:RF_comp_mass_rel_imp}
\end{figure*}

\begin{figure*}
\includegraphics[width=\columnwidth]{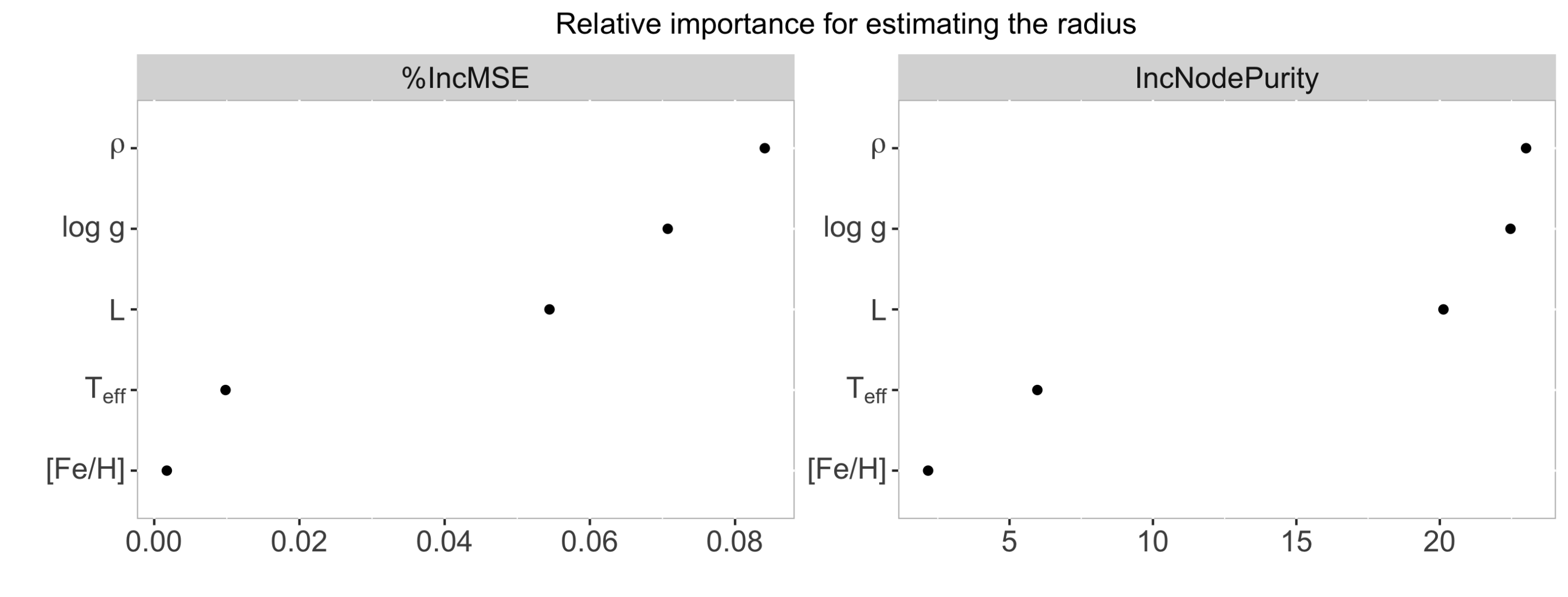}
\caption{Relative importance of the different independent variables in the radius estimations obtained using the Random Forest model. See text for details.}
 \label{fig:RF_comp_radius_rel_imp}
\end{figure*}

We have also tested using machine learning techniques to obtain the best-fitting regressions. Using the complete sample for training a Random Forest model \citep{RF_cite} we obtain an Out-Of-Bag (OOB) mean of squared residual of 0.0043 for estimating M and 0.003 for estimating R, and a percentage of the variance explained by the model of 85.58\,\% and 98.29\,\% for M and R respectively. In Figs. \ref{fig:RF_comp_mass_rel_imp} and \ref{fig:RF_comp_radius_rel_imp} we show the relative importances of the independent variables in the RF regression model for the mass and radius respectively. "$\%$IncMSE" is the increase in "MSE" (Mean Squared Error) of the OOB predictions as a result of variable j being permuted (values randomly shuffled). The higher number, the more important the independent variable. On the other hand, IncNodePurity relates to the variables for which best splits can be chosen in terms of MSE function. More useful variables achieve higher increases in node purities, that is those where you can find a split which has a high inter-node 'variance' and a small intra-node 'variance'. In fact, both plots previde similar but complementary information. In Fig. \ref{fig:RF_comp_mass_rel_imp} we can see that the three variables with the larger values (importance) for the estimation of the mass are $L$, $T_{\rm eff}$, and $\rho$. On the other hand, Fig. \ref{fig:RF_comp_radius_rel_imp} is for the radius and the three variables with larger importance are $\rho$, ${\rm log}g$ and $L$. In both cases these three variables are somehow clustered and clearly different from the other two. Stellar metallicity is always the less important independent variable.

\begin{figure*}
\includegraphics[width=\columnwidth]{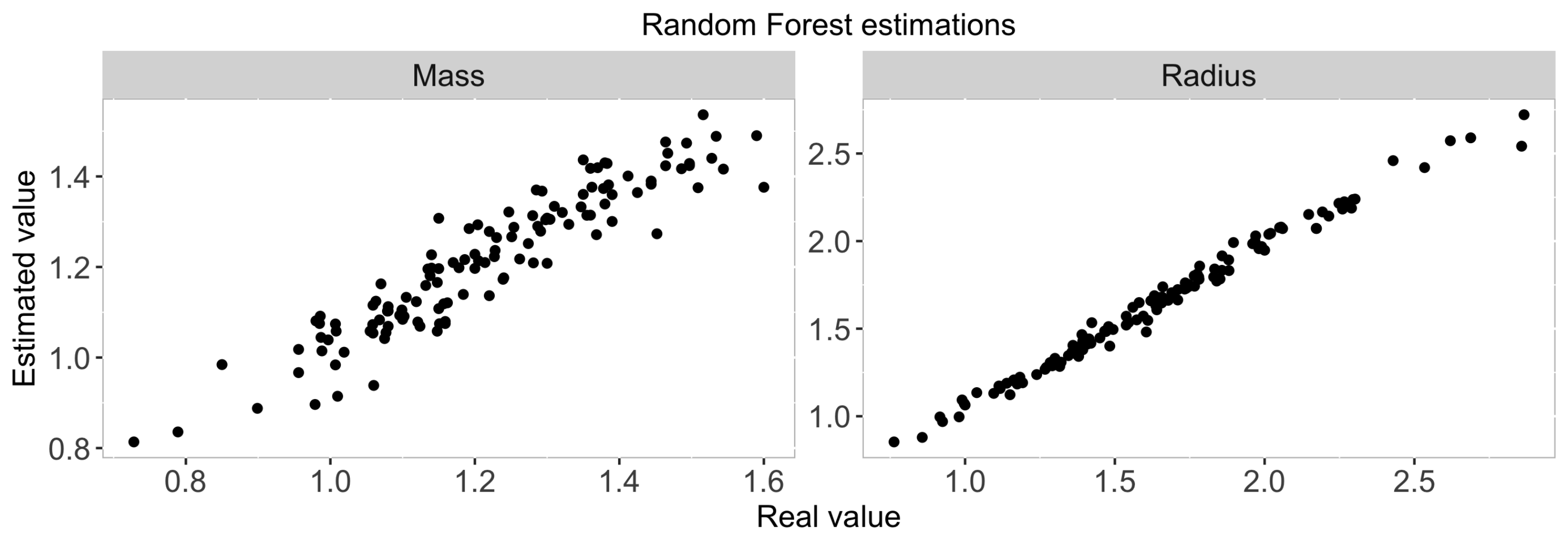}
\caption{Comparison of the mass (left panel) and radius (right panel) estimations obtained using the Random Forest model and the real testing values, where ``Real'' means the value obtained using the techniques described in Section \ref{sec:data_sample}, that is, Asteroseismology, Eclipsing binaries, and Interferometry.}
 \label{fig:RF_comp}
\end{figure*}

\begin{figure*}
\includegraphics[width=\columnwidth]{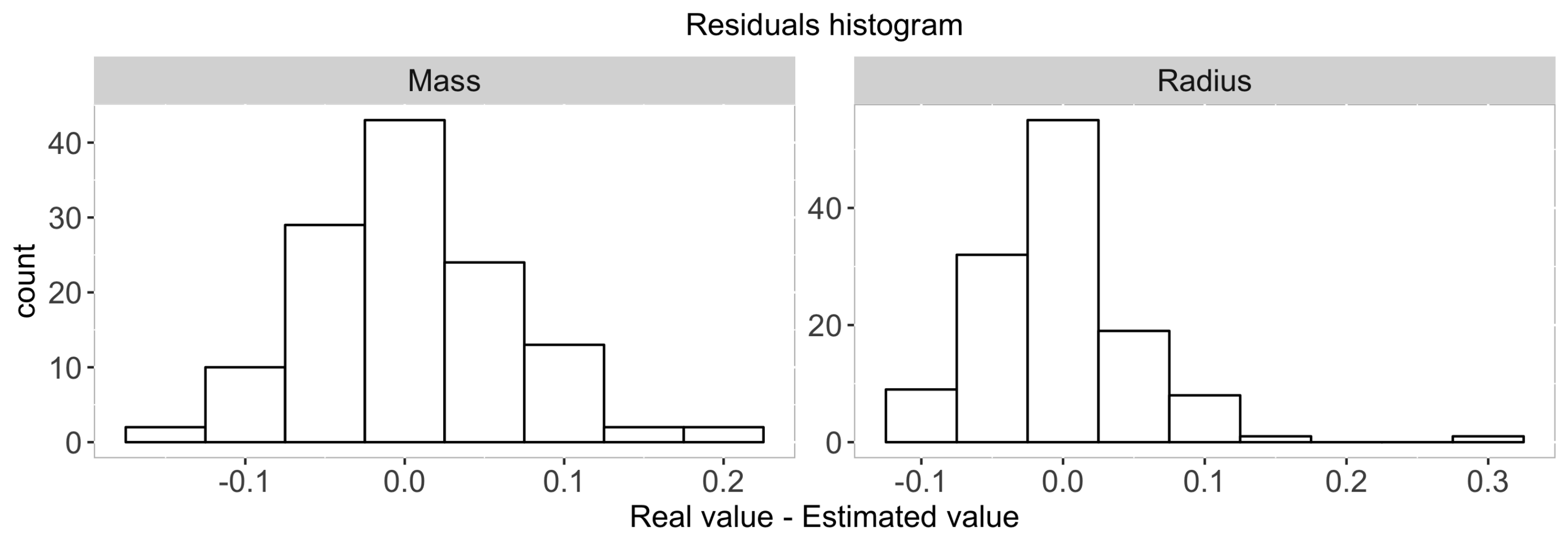}
\caption{Histograms with the residuals of the mass estimations (left panel) and radius estimations (right panel) obtained using the Random Forest model and the real testing values.}
\label{fig:RF_comp_res}
\end{figure*}

In addition, and to illustrate the application of this RF model for estimating masses and radii, we have trained an new Random Forest model using all the independent variables available on 70$\%$ of the MS stars in our sample, using the remaining 30$\%$ as the control group. This split into train and control groups is different from that used for the regressions in the previous sections. In the case of the regressions, the split into train and control groups depends on the uncertainties of the variables involved. In the case of this Random Forest model test, as uncertainties don't play any role, we directly split the complete sample randomly. The comparison of the estimated values and "Real" values for the mass and radius of the testing sample are shown in Fig. \ref{fig:RF_comp} (where "Real" means the values provided by the techniques described in Section \ref{sec:data_sample}, that is, Asteroseismology, Eclipsing binaries, and Interferometry). The implied accuracy is remarkable. Histograms with the residuals of these estimates are shown in Fig \ref{fig:RF_comp_res}. The mean squared residuals of both distributions on the control group are 0.0036 and 0.0026 for M and R respectively, similar to those obtained for the RF model trained with the complete sampling, and the relative accuracies obtained (following the definition in Eq. \ref{eq:acc_def}) are 4.7$\%$ for the mass and 3.3$\%$ for the radius. The Random Forest model evidently provides a very efficient and accurate way for obtaining regression models to estimate the mass and/or the radius. The accuracies reached with this model are similar or better to those obtained with our GLSME models.


\subsection{Comparison with other relations in the literature}

We have compared our results with some of the most recent and popular relations in the literature, in particular, those of \citet{Torres10,Gafeira12}, and \citet{2007MNRAS.382.1073M}. We have not compared with the relations of \citet{Eker14} because they use the luminosity as the dependent variable and the mass as the independent variable, making it impossible to obtain a reliable comparison with our results.

\citet{Torres10} provided one relation for the stellar mass and another for the stellar radius, in the form $f(X,X^2,X^3,{\rm log}^2g,{\rm log}^3g,{\rm [Fe/H]})$, where $X={\rm log}T_{\rm eff} - 4.1$. These relations are comparable to those we present, i.e., those in the form ${\rm log}M\, \mathrm{or}\, {\rm log}R \backsim T_{\rm eff} + {\rm log}g + {\rm [Fe/H]}$. Using the control group of these relations to estimate the relative accuracy and precision obtained using the Torres' equations, we have reached, for the mass, an Acc of 7.37\% and a Prec of 52.86\%. Compared with the overall Acc of 7.54\% and Prec of 3.43\% in table \ref{table:result_r2_acc_prec}, we find that both relations estimate the stellar mass with a good (and similar) accuracy but the precision in the Torres' formula is much deteriorated mainly due to the large number of dimensions. In the case of the radius, Torres' equations give an Acc of 3.64\% and a Prec of 36.02\%, to be compared with our overall Acc of 2.97\% and Prec of 2.73\%. Again, similar accuracies and very different precisions. Therefore, the main difference between Torres' relations and ours is the number of independent variables. The precision achieved, taking into account only the coefficient errors, is favorable to the expression with the lower number of dimensions. And in practice the final precision (when the uncertainties of the inputs are taken into account) gets worse when the numner of dimensions of the relations increases. That is, since Torres' relations involve six variables and ours only three, in terms of precision our relations are preferred for obtaining similar accuracies.

\citet{Gafeira12} provided three relations for the stellar mass. One is a function of ${\rm log}L$, ${\rm log}^2L$ and ${\rm log}^3L$, another adds [Fe/H], [Fe/H]$^2$ and [Fe/H]$^3$ to the previous relation, while a third one adds the stellar age to the second relation. This third relation is not really useful since the stellar age is not known, in general, with good precision (and the accuracy is as unknown). Therefore, we have compared the estimations of the two first relations with ours.

The first relation must be compared with our ${\rm log}M \backsim {\rm log}L$ relation. Their relation, when compared to our control group, provides an Acc of 18.45\% and a Prec of 12.90\%. These values must be compared with our overall Acc of 10.80\% and Prec of 0.13\%. The second relation provides an Acc of 10.43\% and a Prec of 9.87\%. This must be compared with our relation ${\rm log}M \backsim {\rm log}L+{\rm [Fe/H]}$, which gives an overall Acc of 9.91\% and Prec of 0.88\%. The main differences can be understood by the fact that Gafeira's expressions, again, have a larger number of dimensions compared with ours, with the precision deterioration it implies, and they have obtained their relations using only 26 stars.

Finally, we have also compared the $M=f({\rm log}L,{\rm log}^2L)$ and $M=f({\rm log}T_{\rm eff},{\rm log}^2T_{\rm eff},{\rm log}^3T_{\rm eff},{\rm log}^4T_{\rm eff})$ of \citet{2007MNRAS.382.1073M} with our ${\rm log}M \backsim {\rm log}L$ and $M \backsim T_{\rm eff}$ relations, respectively. The first relation of \citet{2007MNRAS.382.1073M} provides an Acc of 11.24\%. This accuracy compares with our overall Acc of 8.29\%. \citet{2007MNRAS.382.1073M} do not provide any errors for their coefficients, and as such we cannot estimate the relative precision of their expressions. The second relation gives an unexpectedly large Acc of 426.91\% (compared to our Acc of 10.08\%). We have tried to reproduce both of Malkov's relations with our data, and in the case of $M=f({\rm log}L,{\rm log}^2L)$ we find similar coefficients, but in the case of $M=f({\rm log}T_{\rm eff},{\rm log}^2T_{\rm eff},{\rm log}^3T_{\rm eff},{\rm log}^4T_{\rm eff})$ we cannot reproduce their results.

\section{Exoplanet host stars}
%
%
Owing to the observational techniques that are used to discover exoplanets, their characterisation is linked to an accurate knowledge of the host star mass and/or radius. At present, only a comparatively small number of planet-hosting stars have been characterised by one of the three source techniques considered by us. Therefore, stellar masses and radii must be estimated sometimes using alternative methods.

To illustrate the impact of using our derived relations, we have applied them to a subset of our stellar sample that comprises 61 planet-hosting stars. In Table \ref{table:result_r2_acc_prec} we display two additional columns: ``Acc. plan'' and ``Prec. plan'', representing the relative accuracy and precision obtained using only stars harboring planets. As expected, these accuracies and precisions are similar to those obtained for the control group.

\section{Summary and future}

In this work, we have taken advantage of the most recent accurate stellar characterizations carried out using asteroseismology, eclipsing binaries and interferometry to evaluate a comprehensive set of empirical relations for the estimation of stellar masses and radii. We have gathered a total of 934 stars -- of which almost two-thirds are on the Main Sequence -- that are characterized with different levels of precision, most of them having estimates of $M$, $R$, $T_{\rm eff}$, $L$, $g$, $\rho$ and [Fe/H]. We have deliberately used a heterogeneous sample (in terms of characterizing techniques and spectroscopic types) to reduce the influence of possible biases coming from the observation, reduction, and analysis methods used to obtain the stellar parameters.

We have studied a total of 576 linear combinations of $T_{\rm eff}$, $L$, $g$, $\rho$ and [Fe/H] (and their logarithms) to be used as independent variables to estimate $M$ or $R$. We have used an error-in-variables regression algorithm (Generalized Least Squares with Measurement Error, GLSME) for a realistic estimation of the regression coefficient's uncertainties. For every combination, we have selected the subset of stars with the lowest uncertainties and applied the GLSME algorithm to them, using the remaining stars as controls. We have used the $R^2$ statistic and the relative accuracy and precision over different control groups to select the best relations over these 576 combinations.

We present a total of 38 new or revised relations, all of which have an $R^2>0.85$ (84$\%$ have $R^2>0.9$); a relative accuracy better than 10$\%$ (aside from three cases); and a relative precision better than 7.5$\%$ (aside from one case). In general, the addition of more dimensions to the relations improves R-squared and the Accuracy, and Precision, deteriorates. Expressions with 2 or 3 dimensions are those with a most compensated balance among R-Squared, Accuracy and Precision. In any case, the particular choosing of a certain relation must be evaluated at each particular case. A subsample of 61 stars in our sample that are planet hosts returns results having similar precision and accuracy to the bulk sample.


We have verified that the use of the standard linear regression provides similar results but with levels of returned precision worst in general than using and error-in-variables model. We have also compared the accuracy and precision obtained using our relations to those given by similar relations in the literature. The various relations provide very similar results, with sometimes better accuracies and precisions returned using our relations.  Finally, we have trained a Random Forest model, which uses machine learning techniques, to estimate $M$ and $R$. This model provides slightly better accuracies when all the variables are taken into account.

In the near future we will focus on the completion of the sampling where it has statistical weaknesses and on obtaining relations suitable for a physical interpretation in terms of their comparison with stellar structure and evolution theories and models.

In sum, this paper serves to provide a revision and extension of empirical relations for the estimation of stellar masses and radii. Finally, we have developed a R package for the estimation of stellar masses and radii using all the tools presented in this work. The package can be found for downloading at https://www.thot-stellar-dating.space/thot-tools/

\acknowledgments

AM acknowledges funding from the European Union's Horizon 2020 research and innovation program under the Marie Sklodowska-Curie grant agreement No 749962 (project THOT). GRD and WJC acknowledge support from the UK Science and Technology Facilities Council (STFC).
Funding for the Stellar Astrophysics Centre is provided by The Danish National Research Foundation (Grant agreement no.: DNRF106). This publication makes use of VOSA, developed under the Spanish Virtual Observatory project supported from the Spanish MICINN through grant AyA2011-24052.

\vspace{5mm}

\software{R version 3.3.1 \citep{R}, RStudio Version 1.0.143, and the R libraries dplyr 0.5.0 \citep{dplyr2016}, xlsx 0.5.7 \citep{xlsx2014}, foreach 1.4.3 \citep{foreach2015}, corrplot 0.86 \citep{corrplot}, GLSME 1.0.3 \citep{GLSME}, parallel \citep{R}, RcppParallel \citep{RcppParallel}, and RandomForest 4.6-12 \citep{RF}}







\begin{longrotatetable}
\begin{deluxetable*}{ccccccccccccc}
\tablecaption{Coefficients and their errors of the selected relations obtained using a GLSME algorithm. \label{table:coefs}
}
\tablewidth{700pt}
\tabletypesize{\scriptsize}
\tablehead{
\colhead{Relation} & \colhead{$a$} & \colhead{$e_a$} & \colhead{$T_{\rm eff}$} & \colhead{$e_{T_{\rm eff}}$} & \colhead{$L$} & \colhead{$e_L$} & \colhead{$g$} & \colhead{$e_g$} & \colhead{$\rho$} & \colhead{$e_\rho$} & \colhead{[Fe/H]} & \colhead{$e_{\rm [Fe/H]}$}}
\startdata
$M = f(T_{\rm eff})$ & $-9.64(-1)$ & $4(-3)$ &  $3.475(-4)$ & $6(-7)$ &        &       &          &       &         &       &        &       \\ 
  ${\rm log}M = f({\rm log}L)$ & $-8(-4)$ & $4(-4)$ &         &       & $2.227(-1)$ & $5(-4)$ &          &       &         &       &        &       \\ 
  $R = f({\rm log}L)$ &  $1.0733$ & $7(-4)$ &         &       & $8.450(-1)$ & $1.3(-4)$ &          &       &         &       &        &       \\ 
  ${\rm log}R = f({\rm log}g)$ &  $3.166(0)$ & $4(-3)$ &         &       &        &       & $-7.122(-1)$ & $9(-4)$ &         &       &        &       \\ 
  ${\rm log}R = f({\rm log}\rho)$ & $-6.1(-3)$ & $6(-4)$ &         &       &        &       &          &       & $-3.99(-1)$ & $1.1(-3)$ &        &       \\ 
  $M = f(T_{\rm eff}+{\rm [Fe/H]})$ & $-9.27(-1)$ & $7(-3)$ &  $3.461(-4)$ & $1.3(-6)$ &        &       &          &       &         &       & $2.18(-1)$ & $3(-3)$ \\ 
  ${\rm log}M = f({\rm log}L+{\rm [Fe/H]})$ &  $1.4(-3)$ & $5(-4)$ &         &       & $2.102(-1)$ & $8(-4)$ &          &       &         &       & $5.7(-2)$ & $1.7(-3)$ \\ 
  ${\rm log}R = f({\rm log}L+{\rm [Fe/H]})$ &  $1.96(-2)$ & $4(-4)$ &         &       & $3.433(-1)$ & $8(-4)$ &          &       &         &       & $4.2(-2)$ & $1.4(-3)$ \\ 
  ${\rm log}R = f({\rm log}g+{\rm [Fe/H]})$ &  $2.871$ & $6(-3)$ &         &       &        &       & $-6.468(-1)$ & $1.4(-3)$ &         &       & $5.57(-2)$ & $1.1(-3)$ \\  
  ${\rm log}R = f({\rm log}\rho+{\rm [Fe/H]})$ & $-4.8(-3)$ & $7(-4)$ &         &       &        &       &          &       & $-3.972(-1)$ & $1.3(-3)$ & $2.62(-2)$ & $2.0(-3)$ \\ 
  ${\rm log}M = f(T_{\rm eff}+{\rm log}L)$ & $-1.19(-1)$ & $3(-3)$ &  $2.14(-5)$ & $5(-7)$ & $1.837(-1)$ & $1.1(-3)$ &          &       &         &       &        &       \\ 
  ${\rm log}R = f(T_{\rm eff}+{\rm log}L)$ &  $5.940(-1)$ & $2.2(-3)$ &  $-1.024(-4)$ & $4(-7)$ & $4.591(-1)$ & $8(-4)$ &          &       &         &       &        &       \\ 
  $M = f(T_{\rm eff}+{\rm log}g)$ & $1.342(0)$ & $2.0(-2)$ &  $2.986(-4)$ & $8(-7)$ &        &       & $-4.64(-1)$ & $4(-3)$ &         &       &        &       \\ 
  ${\rm log}R = f(T_{\rm eff}+{\rm log}g)$ &  $2.444(0)$ & $5(-3)$ &  $4.132(-5)$ & $1.8(-7)$ &        &       & $-6.0525(-1)$ & $1.0(-4)$ &         &       &        &       \\ 
  $M = f(T_{\rm eff}+{\rm log}\rho)$ & $-7.23(-1)$ & $1.7(-2)$ &  $2.92(-4)$ & $3(-6)$ &        &       &          &       & $-3.29(-1)$ & $7(-3)$ &        &       \\ 
  ${\rm log}R  = f(T_{\rm eff}+{\rm log}\rho)$ &  $-1.91(-1) $ & $5(-3)$ &  $3.18(-5)$ & $8(-7)$ &        &       &          &       & $-3.803(-1)$ & $1.2(-3)$ &        &       \\ 
  $M = f({\rm log}L+{\rm log}g)$ &  $-7.5(-1)$ & $3(-2)$ &         &       & $7.60(-1)$ & $3(-3)$ &  $3.96(-1)$ & $7(-3)$ &         &       &        &       \\ 
  ${\rm log}R = f({\rm log}L+{\rm log}g)$ &  $1.915(0)$ & $8(-3)$ &         &       & $1.325(-1)$ & $7(-4)$ & $-4.318(-1)$ & $1.7(-3)$ &         &       &        &       \\ 
  ${\rm log}M = f({\rm log}L+{\rm log}\rho)$ & $-1.70(-2)$ & $1.6(-3)$ &         &       & $2.890(-1)$ & $2.1(-3)$ &          &       &  $8.3(-2)$ & $3(-3)$ &        &       \\ 
  ${\rm log}R = f({\rm log}L+{\rm log}\rho)$ & $-5.9(-3)$ & $6(-4)$ &         &       & $1.08(-1)$ & $3(-3)$ &          &       & $-2.93(-1)$ & $3(-3)$ &        &       \\  
  $M = f(T_{\rm eff}+L+{\rm [Fe/H]})$ & $-3.16(-1)$ & $1.0(-2)$ &  $2.289(-4)$ & $1.8(-6)$ & $3.88(-2)$ & $5(-4)$ &          &       &         &       & $1.31(-1)$ & $4(-3)$ \\ 
  ${\rm log}R = f(T_{\rm eff}+{\rm log}L+{\rm [Fe/H]})$ &  $6.64(-1)$ & $6(-3)$ & $-1.141(-4)$ & $1.0(-6)$ & $4.617(-1)$ & $1.3(-3)$ &          &       &         &       & $9.2(-3)$ & $1.4(-3)$ \\ 
  ${\rm log}M = f(T_{\rm eff}+{\rm log}g+{\rm [Fe/H]})$ & $4.58(-1)$ & $1.1(-2)$ &  $8.43(-5)$ & $5(-7)$ &        &       & $-2.1244(-1)$ & $1.4(-4)$ &         &       & $8.64(-2)$ & $2.2(-3)$ \\ 
  ${\rm log}R = f(T_{\rm eff}+{\rm log}g+{\rm [Fe/H]})$ &  $2.325(0)$ & $9(-3)$ &  $5.04(-5)$ & $6(-7)$ &        &       & $-5.894(-1)$ & $1.6(-3)$ &         &       & $5.88(-2)$ & $1.1(-3)$ \\ 
  $M = f(T_{\rm eff}+{\rm log}\rho+{\rm [Fe/H]})$ & $-5.5(-1)$ & $5(-2)$ &  $2.63(-4)$ & $8(-6)$ &        &       &          &       & $-3.59(-1)$ & $9(-3)$ & $7.4(-2)$ & $1.2(-2)$ \\ 
  ${\rm log}R  = f(T_{\rm eff}+{\rm log}\rho+{\rm [Fe/H]})$ &  $-2.48(-1)$ & $8(-3)$ &  $4.23(-5)$ & $1.4(-6)$ &        &       &          &       & $-3.702(-1)$ & $1.5(-3)$ & $3.24(-2)$ & $2.0(-3)$ \\ 
  ${\rm log}M = f({\rm log}L+g+{\rm [Fe/H]})$ &  $-1.08(-1)$ & $3(-3)$ &         &       & $2.850(-1)$ & $2.1(-3)$ &  $4.25(-6)$ & $1.1(-7)$ &         &       & $5.02(-2)$ & $1.7(-3)$ \\ 
  ${\rm log}R = f({\rm log}L+g+{\rm [Fe/H]})$ &  $2.27(-1)$ & $3(-3)$ &         &       & $1.757(-1)$ & $2.1(-3)$ & $-8.10(-6)$ & $9(-8)$ &         &       & $1.92(-2)$ & $1.5(-3)$ \\ 
  $M = f({\rm log}L+\rho+{\rm [Fe/H]})$ &  $6.60(-1)$ & $1.5(-2)$ &         &       & $9.09(-1)$ & $1.9(-2)$ &          &       &  $3.21(-1)$ & $1.6(-2)$ & $1.81(-1)$ & $1.4(-2)$ \\ 
  ${\rm log}R = f({\rm log}L+{\rm log}\rho+{\rm [Fe/H]})$ &  $-2.8(-3)$ & $7(-4)$ &         &       & $1.18(-1)$ & $4(-3)$ &          &       & $-2.79(-1)$ & $4(-3)$ & $2.59(-2)$ & $2.0(-3)$ \\ 
  $M = f(T_{\rm eff}+L+{\rm log}g)$ &  $1.713(0)$ & $2.4(-2)$ & $2.459(-4)$ & $1.1(-6)$ & $6.52(-3)$ & $1.4(-4)$ & $-4.77(-1)$ & $5(-3)$ &         &       &        &       \\ 
  ${\rm log}R = f(T_{\rm eff}+L+{\rm log}g)$ &  $2.473(0)$ & $6(-3)$ &  $4.44(-5)$ & $3(-7)$ & $-4.37(-4)$ & $2.2(-5)$ & $-6.16(-1)$ & $1.2(-3)$ &         &       &        &       \\ 
  $M  = f(T_{\rm eff}+{\rm log}L+\rho)$ &  $1.3(-1)$ & $3(-2)$ &  $1.09(-4)$ & $6(-6)$ & $7.02(-1)$ & $2.1(-2)$ &          &       &  $2.06(-1)$ & $1.3(-2)$ &        &       \\ 
  $R = f(T_{\rm eff}+L+{\rm log}\rho)$ &  $1.38(0)$ & $2.3(-2)$ & $-7.3(-5)$ & $4(-6)$ & $3.51(-2)$ & $8(-4)$ &          &       & $-1.077(0)$ & $5(-3)$ &        &       \\ 
  $M = f(T_{\rm eff}+L+{\rm log}g+{\rm [Fe/H]})$ &  $4.9(-1)$ & $4(-2)$ & $2.240(-4)$ & $1.7(-6)$ & $3.130(-2)$ & $1.4(-4)$ & $-1.75(-1)$ & $5(-3)$ &         &       & $1.24(-1)$ & $5(-3)$ \\ 
  ${\rm log}R = f(T_{\rm eff}+{\rm log}L+g+{\rm [Fe/H]})$ &  $6.34(-1)$ & $7(-3)$ & $-1.026(-4)$ & $1.5(-6)$ & $4.21(-1)$ & $4(-3)$ & $-1.37(-6)$ & $1.4(-7)$ &         &       & $9.5(-3)$ & $1.5(-3)$ \\ 
  $M  = f(T_{\rm eff}+{\rm log}L+\rho+{\rm [Fe/H]})$ & $2.7(-1)$ & $6(-2)$ &  $7.1(-5)$ & $1.1(-5)$ & $8.32(-1)$ & $2.3(-2)$ &          &       &  $3.07(-1)$ & $1.6(-2)$ & $1.77(-1)$ & $1.4(-2)$ \\ 
  ${\rm log}R = f(T_{\rm eff}+L+{\rm log}\rho+{\rm [Fe/H]})$ &  $-2.24(-1)$ & $8(-3)$ & $3.78(-5)$ & $1.4(-6)$ & $3.9(-3)$ & $4(-4)$ &          &       & $-3.48(-1)$ & $3(-3)$ & $3.72(-2)$ & $2.1(-3)$ \\ 
\enddata
\end{deluxetable*}
\end{longrotatetable}

\begin{longrotatetable}
\begin{deluxetable*}{ccccccccccc}
\tablecaption{Validity ranges of each relation. \label{table:ranges}
}
\tablehead{
\colhead{Rel} & \colhead{$T_{\rm eff,min}$} & \colhead{$T_{\rm eff,max}$} & \colhead{${\rm log}L_{\rm min}$} & \colhead{${\rm log}L_{\rm max}$} & \colhead{${\rm log}g_{\rm min}$} & \colhead{${\rm log}g_{\rm max}$} & \colhead{$\rho_{\rm min}$} & \colhead{$\rho_{\rm max}$} & \colhead{[Fe/H]$_{\rm min}$} & \colhead{[Fe/H]$_{\rm max}$}} 
\startdata
$M \backsim T_{\rm eff}$ & 4780 & 11100 &             &             &             &             &             &             &             &             \\ 
${\rm log}M \backsim {\rm log}L$ &              &             &    -0.717 &     2.010 &             &             &             &             &             &             \\ 
$R \backsim {\rm log}L$ &              &             &    -0.717 &     2.010 &             &             &             &             &             &             \\ 
 ${\rm log}R \backsim {\rm log}g$ &              &             &             &             &     3.485 &     4.653 &             &             &             &             \\ 
 ${\rm log}R \backsim {\rm log}\rho$ &              &             &             &             &             &             &     0.063 &     1.774 &             &             \\ 
 $M \backsim T_{\rm eff} + {\rm [Fe/H]}$ &   4780 & 10025 &             &             &             &             &             &             &    -0.920 &     0.380 \\ 
 ${\rm log}M \backsim {\rm log}L + {\rm [Fe/H]}$ &              &             &    -0.717 &     1.363 &             &             &             &             &    -0.920 &     0.380 \\ 
 ${\rm log}R \backsim {\rm log}L + {\rm [Fe/H]}$ &              &             &    -0.717 &     1.363 &             &             &             &             &    -0.920 &     0.380 \\ 
 ${\rm log}R \backsim {\rm log}g + {\rm [Fe/H]}$ &              &             &             &             &     3.701 &     4.653 &             &             &    -0.920 &     0.380 \\ 
 ${\rm log}R \backsim {\rm log}\rho + {\rm [Fe/H]}$ &              &             &             &             &             &             &     0.064 &     1.774 &    -0.920 &     0.380 \\ 
 ${\rm log}M \backsim T_{\rm eff} + {\rm log}L$ &   4780 & 10990 &    -0.717 &     2.010 &             &             &             &             &             &             \\ 
 ${\rm log}R \backsim T_{\rm eff} + {\rm log}L$ &   4780 & 10990 &    -0.717 &     2.010 &             &             &             &             &             &             \\ 
 $M \backsim T_{\rm eff} + {\rm log}g$ &   4780 & 11100 &             &             &     3.485 &     4.653 &             &             &             &             \\ 
 ${\rm log}R \backsim T_{\rm eff} + {\rm log}g$ &   4780 & 12000 &             &             &     3.485 &     4.653 &             &             &             &             \\ 
 $M \backsim T_{\rm eff} + {\rm log}\rho$ &   5046 &  9950 &             &             &             &             &     0.064 &     1.774 &             &             \\ 
 ${\rm log}R  \backsim T_{\rm eff} + {\rm log}\rho$ &   5046 &  9950 &             &             &             &             &     0.064 &     1.774 &             &             \\ 
  $M \backsim {\rm log}L + {\rm log}g$ &             &             &    -0.717 &     2.010 &     3.580 &     4.653 &             &             &             &             \\ 
  ${\rm log}R \backsim {\rm log}L + {\rm log}g$ &             &             &    -0.717 &     2.010 &     3.545 &     4.653 &             &             &             &             \\ 
 ${\rm log}M \backsim {\rm log}L + {\rm log}\rho$ &              &             &    -0.467 &     1.480 &             &             &     0.064 &     1.774 &             &             \\ 
 ${\rm log}R \backsim {\rm log}L + {\rm log}\rho$ &              &             &    -0.467 &     1.480 &             &             &     0.064 &     1.774 &             &             \\ 
 $M \backsim T_{\rm eff} + L + {\rm [Fe/H]}$ &   4780 &  9000 &    -0.717 &     1.363 &             &             &             &             &    -0.920 &     0.380 \\ 
 ${\rm log}R \backsim T_{\rm eff} + {\rm log}L + {\rm [Fe/H]}$ &   4780 &  9000 &    -0.717 &     1.363 &             &             &             &             &    -0.920 &     0.380 \\ 
 ${\rm log}M \backsim T_{\rm eff} + {\rm log}g + {\rm [Fe/H]}$ &   4780 & 10025 &             &             &     3.701 &     4.653 &             &             &    -0.920 &     0.380 \\ 
 ${\rm log}R \backsim T_{\rm eff} + {\rm log}g + {\rm [Fe/H]}$ &   4780 & 10025 &             &             &     3.701 &     4.653 &             &             &    -0.920 &     0.380 \\ 
 $M \backsim T_{\rm eff} + {\rm log}\rho + {\rm [Fe/H]}$ &   5046 &  7030 &             &             &             &             &     0.064 &     1.774 &    -0.920 &     0.380 \\ 
 ${\rm log}R  \backsim T_{\rm eff} + {\rm log}\rho + {\rm [Fe/H]}$ &   5046 &  7030 &             &             &             &             &     0.064 &     1.774 &    -0.920 &     0.380 \\ 
 ${\rm log}M \backsim {\rm log}L + g + {\rm [Fe/H]}$ &              &             &    -0.717 &     1.167 &     3.701 &     4.653 &             &             &    -0.920 &     0.380 \\ 
 ${\rm log}R \backsim {\rm log}L + g + {\rm [Fe/H]}$ &              &             &    -0.717 &     1.167 &     3.701 &     4.653 &             &             &    -0.920 &     0.380 \\ 
 $M \backsim {\rm log}L + \rho + {\rm [Fe/H]}$ &              &             &    -0.467 &     1.153 &             &             &     0.064 &     1.774 &    -0.920 &     0.380 \\ 
 ${\rm log}R \backsim {\rm log}L + {\rm log}\rho + {\rm [Fe/H]}$ &              &             &    -0.467 &     1.153 &             &             &     0.064 &     1.774 &    -0.920 &     0.380 \\ 
 $M \backsim T_{\rm eff} + L + {\rm log}g$ &   4780 & 10700 &    -0.717 &     2.010 &     3.580 &     4.653 &             &             &             &             \\ 
 ${\rm log}R \backsim T_{\rm eff} + L + {\rm log}g$ &   4780 & 10700 &    -0.717 &     2.010 &     3.545 &     4.653 &             &             &             &             \\ 
  $M  \backsim T_{\rm eff} + {\rm log}L + \rho$ &  5046 &  9950 &    -0.467 &     1.480 &             &             &     0.064 &     1.774 &             &             \\ 
 $R \backsim T_{\rm eff} + L + {\rm log}\rho$ &   5046 &  9950 &    -0.467 &     1.480 &             &             &     0.064 &     1.774 &             &             \\ 
 $M \backsim T_{\rm eff} + L + {\rm log}g + {\rm [Fe/H]}$ &   4780 &  9000 &    -0.717 &     1.167 &     3.701 &     4.653 &             &             &    -0.920 &     0.380 \\ 
 ${\rm log}R \backsim T_{\rm eff} + {\rm log}L + g + {\rm [Fe/H]}$ &   4780 &  9000 &    -0.717 &     1.167 &     3.701 &     4.653 &             &             &    -0.920 &     0.380 \\ 
 $M  \backsim T_{\rm eff} + {\rm log}L + \rho + {\rm [Fe/H]}$ &   5046 &  7030 &    -0.467 &     1.153 &             &             &     0.064 &     1.774 &    -0.920 &     0.380 \\ 
 ${\rm log}R \backsim T_{\rm eff} + L + {\rm log}\rho + {\rm [Fe/H]}$ &   5046 &  7030 &    -0.467 &     1.153 &             &             &     0.064 &     1.774 &    -0.920 &     0.380 \\ 
\enddata
\end{deluxetable*}
\end{longrotatetable}

\appendix





\section{Cross-correlation between the independent variables}
\label{sec:cross-cor}

\begin{figure*}
\includegraphics[width=\columnwidth]{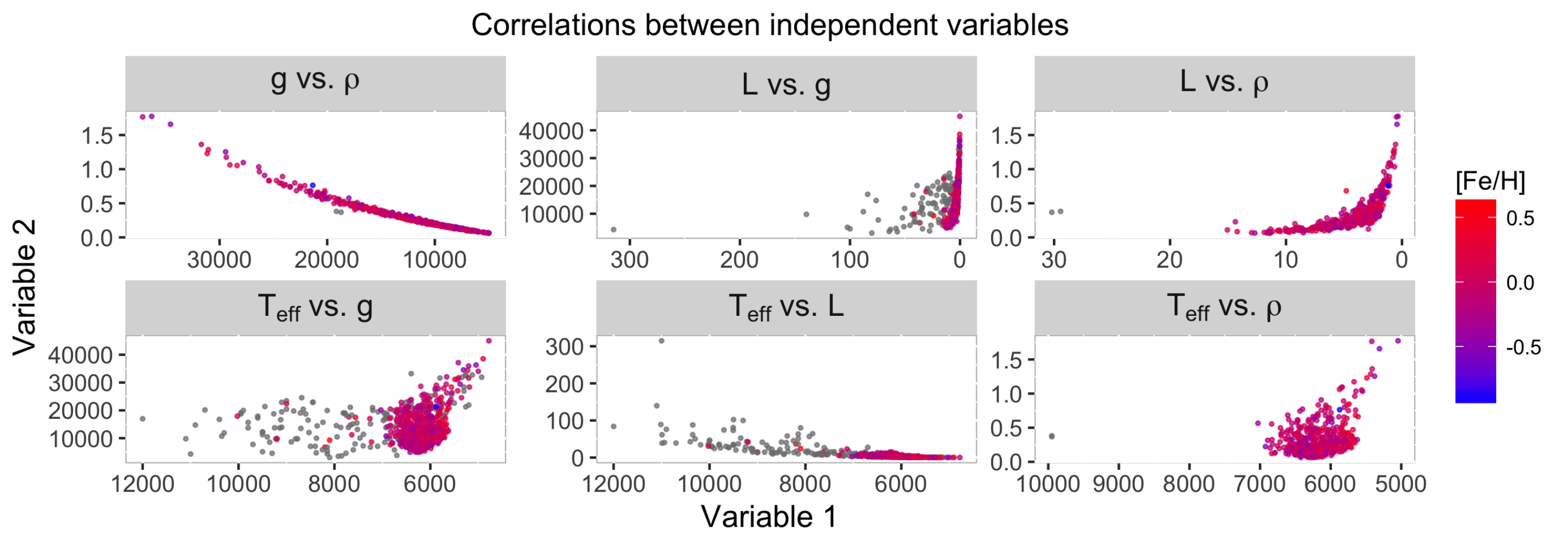}
\caption{Scatter plots of the different independent variables. First variable is represent at x-axes, the second one at y-axes. The metallicity is shown in color scale. Grey dots are those with unknown metallicity.}
\label{fig:cross_cor}
\end{figure*}

In section \ref{sec:comb_var} we analyzed the cross-correlations between the independent variables of our study. As a complement to Fig. \ref{fig:correl}, in Fig. \ref{fig:cross_cor} we show the scatter plot of the different pairs of variables. Here we can verify the information provided by the Kendall - $\tau$ coefficient. In general we can see that most of the stars are located in a certain zone or line, something we can regard as a "Main-Sequence" behavior. In any case, all cross-correlations except g vs. $\rho$ and L vs. $\rho$ present a large dispersion, enough for regarding that each variable can provide independent and complementary information. L vs. $\rho$ has a non-linear function-like behavior, with a large spread at the elbow. This spread allows the use of both variables at the same time, since both can provide some complementary information. Finally, g and $\rho$ are clearly correlated.

\end{document}